\documentclass[prd,amsmath,amssymb,aps,superscriptaddress,twocolumn,nofootinbib
]{revtex4-2}

\usepackage{epsfig}
\usepackage{url}
\usepackage{hyperref}

\usepackage[normalem]{ulem}

\usepackage{latexsym}
\usepackage{epsfig}
\usepackage{amsmath}
\usepackage{amssymb}
\usepackage{wasysym}
\usepackage{graphicx}
\usepackage{dcolumn}
\usepackage{float}
\usepackage{verbatim}
\usepackage{enumerate,mdwlist}%lists

\usepackage[titletoc]{appendix}

\usepackage{amsfonts}
\usepackage{fancyvrb} % for "\Verb" macro
\usepackage{tikz} %diagrams
\usetikzlibrary{calc}
\usepackage[export]{adjustbox}
\usepackage[normalem]{ulem}%strikethrough text

\usepackage[inline, final]{showlabels}

\usepackage{listings}
\usepackage{xcolor}
\usepackage{amsmath, orcidlink}

\definecolor{comment_red}{rgb}{0.5, 0, 0}
\makeatletter
\g@addto@macro\bfseries{\boldmath}  % automatic bold math in sections
\makeatother

\newcommand{\ba}{\begin{eqnarray}}
\newcommand{\ea}{\end{eqnarray}}
\definecolor{grey}{rgb}{0.4,0.4,0.4}
\definecolor{dullmagenta}{rgb}{0.4,0,0.4}
\definecolor{darkblue}{rgb}{0,0,0.4}
\definecolor{midblue}{rgb}{0,0,0.5}
\definecolor{midred}{rgb}{0.5,0,0}
\definecolor{orange}{rgb}{1,0.5,0}
\definecolor{lightbrown}{rgb}{0.75,0.5,0.25}
\definecolor{tan}{cmyk}{0.14,0.42,0.56,0}
\definecolor{djunglegreen}{cmyk}{0.99,0,0.52,0}
\definecolor{lightgreen}{rgb}{0,1,0}
\definecolor{olivegreen}{cmyk}{0.64,0,0.95,0.40}
\definecolor{midgreen}{rgb}{0.0,0.675,0.0}
\definecolor{darkgreen}{rgb}{0,0.5,0}
\definecolor{ceruleanblue}{rgb}{0.0, 0.2, 0.7}
\definecolor{burgundy}{rgb}{0.5, 0.0, 0.13}
\definecolor{hvred}{RGB}{186,12,47}

\hypersetup{
    colorlinks=true,
    linkcolor=ceruleanblue,
    filecolor=ceruleanblue,
    urlcolor=midblue,
    citecolor=burgundy,
}

\usepackage{booktabs}

\usepackage[all]{xy} %array diagrams
\usepackage{amsfonts}

\makeatletter
\def\l@subsubsection#1#2{}
\makeatother

\begin{document}

\title{From Stars to Waves: %Non-deterministic Inference of Microlensed Gravitational Waves}
Stochastic Inference of Microlensed Gravitational Waves}
% repeat the \author .. \affiliation  etc. as needed
% \email, \thanks, \homepage, \altaffiliation all apply to the current
% author. Explanatory text should go in the []'s, actual e-mail
% address or url should go in the {}'s for \email and \homepage.
% Please use the appropriate macro foreach each type of information

% \affiliation command applies to all authors since the last
% \affiliation command. The \affiliation command should follow the
% other information
% \affiliation can be followed by \email, \homepage, \thanks as well.
%\author{}
%\email[]{Your e-mail address}
%\homepage[]{Your web page}
%\thanks{}
%\altaffiliation{}
%\affiliation{}
\author{Zhaoqi Su \orcidlink{0000-0003-3651-8373}}
\thanks{Zhaoqi Su and Xikai Shan contributed equally to this work.}
\affiliation{College of Physics and Information Engineering, Fuzhou University, Fuzhou 350108, China}
\affiliation{Department of Automation, Tsinghua University, Beijing 100084, China}

\author{Xikai Shan \orcidlink{0000-0003-3201-061X}}
\thanks{Zhaoqi Su and Xikai Shan contributed equally to this work.}
% \email{xikai_shan@mail.bnu.edu.cn} 
\affiliation{Department of Astronomy, Tsinghua University, Beijing 100084, China}

\author{Zhenwei Lyu \orcidlink{0000-0002-0151-3794}}
\affiliation{Leicester International Institute, Dalian University of Technology, Panjin 124221, China}

\author{Junyao Zhang}
\affiliation{Department of Physics, Tsinghua University, Beijing 100084, China}

\author{Yebin Liu}
\affiliation{Department of Automation, Tsinghua University, Beijing 100084, China}

\author{Shude Mao \orcidlink{0000-0001-8317-2788}}
\affiliation{Department of Astronomy, Westlake University, Hangzhou 310030, Zhejiang Province, China}

\author{Huan Yang \orcidlink{0000-0002-9965-3030}}
\email{hyangdoa@tsinghua.edu.cn} \affiliation{Department of Astronomy, Tsinghua University, Beijing 100084, China}

%Collaboration name if desired (requires use of superscriptaddress
%option in \documentclass). \noaffiliation is required (may also be
%used with the \author command).
%\collaboration can be followed by \email, \homepage, \thanks as well.
%\collaboration{}
%\noaffiliation

\date{\today}

\begin{abstract}
    When strongly lensed gravitational waves pass through the stellar field of a lensing galaxy, they acquire additional phase and amplitude modulations from gravitational microlensing by stars and stellar remnants along the line of sight.
    These microlensed waveforms depend on the masses and positions of thousands or more of the most relevant stars required for numerical convergence of the wave-optics diffraction integral, so that their deterministic reconstruction from the data is computationally prohibitive. 
    We formulate the detection and parameter estimation of such events as a stochastic inference problem
    and propose a solution with the implementation of normalizing flows. As a proof of principle, we use the inferred surrogate microlensing parameters to identify microlensing signatures and show that 10.2\% of microlensed events can be detected at $\ge 3\sigma$ significance with third-generation gravitational-wave detector networks for source redshifts $z_s \le 2$. The correlation between the surrogate parameters and the density of the underlying stellar field further connects microlensing detection to the properties of the lensing galaxy.
    This approach opens the door to probing microlensing effects and the properties of the underlying stellar fields. Beyond microlensed gravitational waves, a similar construction may also be applied to other intrinsically stochastic inference problems, such as detecting post-merger gravitational waves from binary neutron star coalescence and signals from core-collapse supernovae.
\end{abstract}

% insert suggested keywords - APS authors don't need to do this
%\keywords{}

%\maketitle must follow title, authors, abstract, and keywords
\maketitle

\section{Introduction}
% Strongly-lensed gravitational waves (SLGWs) are emerging as novel tracers with unique advantages in cosmological and astrophysical studies~\citep{Li:2018prc, Oguri:2019fix, Hannuksela:2020xor, 2022A&A...659L...5C, Liao:2022gde, PhysRevLett.130.261401, Seo_2024, 2025MNRAS.536.2212P, 2022PhRvD.106b3018G, 2020PhRvD.102l4076G, Singh:2025uvp, Menadeo:2024uoq, Villarrubia-Rojo:2024xcj, Brando:2024inp, Cheung:2024ugg, Savastano:2023spl, Goyal:2023uvm, Tambalo:2022wlm, Tambalo:2022plm}. For instance, millisecond-level time-delay measurements from strongly-lensed binary black holes offer the potential for a more precise estimation of the Hubble parameter~\citep{Liao:2017ioi}, improve the localization precision of binary black hole mergers~\citep{Hannuksela:2020xor, Yu:2020agu} to facilitate the identification of their host galaxy, and also serve as a novel platform for testing general relativity~\citep{Baker:2016reh, Collett:2016dey, Fan:2016swi}.

Strongly lensed gravitational waves (SLGWs) are emerging as novel tracers for cosmology and astrophysics~\citep{Li:2018prc, Oguri:2019fix, Hannuksela:2020xor, 2022A&A...659L...5C, Liao:2022gde, PhysRevLett.130.261401, Seo_2024, 2025MNRAS.536.2212P, 2022PhRvD.106b3018G, 2020PhRvD.102l4076G, Singh:2025uvp, Menadeo:2024uoq, Villarrubia-Rojo:2024xcj, Brando:2024inp, Cheung:2024ugg, Savastano:2023spl, Goyal:2023uvm, Tambalo:2022wlm, Tambalo:2022plm}, with applications including millisecond-level time-delay measurements for Hubble-parameter estimation~\citep{Liao:2017ioi}, improved localization of binary black hole mergers for host-galaxy identification~\citep{Hannuksela:2020xor, Yu:2020agu}, and tests of general relativity~\citep{Baker:2016reh, Collett:2016dey, Fan:2016swi}. In the strong-lensing regime, however, an individual image commonly traverses a stellar field in the lens galaxy. When the gravitational-wave wavelength is comparable to the Schwarzschild radii of stellar-mass compact objects, wave-optics microlensing imprints diffraction and interference features on the waveform. This work focuses on these microlensing signatures: we develop a framework to infer and assess stellar-field microlensing in gravitational-wave data, rather than to identify strong lensing through image multiplets.

This microlensing-induced wave-optics modulation is common for SLGWs that pass through a field of stellar-mass microlenses within the strong-lensing galaxy~\citep{Shan:2022xfx, Shan:2023qvd, Shan:2023ngi, Shan:2024min, Meena:2022unp, Diego:2019lcd, Seo:2025dto}.
\citet{Shan:2024min} found that approximately 90\% of such events have a waveform mismatch larger than $10^{-4}$ between signals with and without stellar microlensing.
If these oscillations are faithfully detected and accurately characterized, they can be used to constrain the mass distribution of the microlenses.
This provides an opportunity to study compact-object populations in stellar microlensing fields, such as stars and stellar remnants along the line of sight.
Traditional quasar-microlensing methods rely on long-term photometric monitoring of strongly lensed quasar images and on statistics of the resulting microlensing variability to constrain properties such as the characteristic microlens mass scale~\citep{Kochanek_2004, Morgan_2006, vernardos2023microlensingstronglylensedquasars}. The GW problem is qualitatively different: a GW event is a short-duration chirp, and the microlensing information is encoded in a single frequency-dependent wave-optics interference pattern rather than in a light curve produced as a source moves across a magnification map over months or years. Therefore, the traditional statistical approach developed for quasar microlensing cannot be transferred directly to the present problem.
% \red{However, such lensed signals with wave-optics features can be severely down-ranked in current matched-filter searches, potentially leading to low detection efficiency \cite{PhysRevD.111.084019}.}

The recent GW event GW231123~\citep{2025arXiv250708219T} was reported with component masses of $137^{+22}_{-17} \mathrm{M}_\odot$ and $103^{+20}_{-52} \mathrm{M}_\odot$ (90\% credible interval).
These features are unusual, particularly because the secondary mass falls within the theoretical pair-instability mass gap (about 60–130$\mathrm{M}_\odot$~\citep{2019ApJ...887...53F, 2020ApJ...902L..36F, 2021ApJ...912L..31W, Hendriks:2023yrw}) and the spins are near the extremal Kerr limit~\citep{PhysRevLett.11.237, PhysRev.174.1559}. These unconventional characteristics could be naturally explained if the event were subject to gravitational lensing.
%falls within the upper mass gap, a feature that lensing could explain.
% For instance, high magnification would reduce the inferred source masses to values below the gap~\citep{2018arXiv180205273B, 2020arXiv200208821B, Broadhurst:2020cvm,2023MNRAS.521.3421B}, while the observed precession could be mimicked by microlensing wave effects in SLGWs~\citep{Liu:2023emk,Shan:2025new}. 
For instance, high magnification can reduce the inferred source masses to values below the gap~\citep{2018arXiv180205273B, 2020arXiv200208821B, Broadhurst:2020cvm,2023MNRAS.521.3421B}. The source-frame mass of a GW event is related to the detector-frame mass by $M_{\rm o} = (1+z) M_{\rm z}$. If an event is strongly lensed, the magnification can lead us to underestimate the luminosity distance to the source. This, in turn, causes us to underestimate the redshift and therefore overestimate the source-frame mass $M_{\rm z}$. For this reason, some studies have suggested that strong lensing may explain GW sources with unusually large inferred masses, especially those in the mass gap. In addition, the observed precession could be mimicked by microlensing wave effects in SLGWs~\citep{Liu:2023emk,Shan:2025new}, since both precession and diffraction/interference can produce similar beat patterns in the waveform.
This highlights the importance of developing tools to study the combined effects of strong lensing and microlensing on SLGWs.

The successful identification of microlensing signatures is a critical prerequisite for exploiting strongly lensed gravitational waves in cosmological and astrophysical studies. Several approaches have been proposed for strongly lensed gravitational wave identification~\citep{Haris:2018vmn, Barsode2024FastWaves, Kim:2020xkm, Goyal:2021hxv, Lo:2021nae, Liu:2020par, Janquart:2021qov, Janquart_2023, Dai:2017huk, Wang:2021kzt, Janquart:2021nus, Ezquiaga:2023xfe, Wright:2025nod, Chakraborty:2024net}, Monte Carlo-type methods to identify isolated microlensing events~\citep{Wright:2021cbn, Nerin_2025, chakraborty2025modelindependentchromaticmicrolensingsearch, PhysRevD.111.084019, 2025arXiv250520996Q}, and template-free methods to recover microlensing-induced distortions~\citep{Dai:2018enj, Zumalacarregui:2024ocb, Shan:2023ngi, Chakraborty:2024mbr}. Dedicated pipelines have also been developed to detect microlensed GW signals, such as Gravelamps~\citep{Wright:2021cbn}, GLoW~\citep{Villarrubia-Rojo:2024xcj}, and Dingo-lensing~\citep{Chan:2025pdf}. In particular, GLoW provides an efficient framework for wave-optics calculations of generic lens configurations, including systems with multiple point-mass lenses. However, existing methods are not designed for stochastic stellar microlensing involving large stellar fields containing thousands of stars. 
%However, these tools focus on simple models, like isolated lenses or point lenses within a galaxy. 
In reality, a GW waveform is lensed by a stellar field containing thousands to millions of stars. Extracting information from all these stars is computationally impossible. Therefore, we need a simpler way to capture this signature. This work tests whether it is feasible to infer a few phenomenological surrogate parameters from the waveform. The present approach is intended to identify microlensing signatures in an individual macroimage rather than to identify strongly lensed multiple-image pairs, and may also be used to infer properties of the stellar field in the future.

%However, notice that the same set of characteristic observables can arise from many different stellar-field realizations, meaning that a waveform containing only the characteristic observables is stochastic. 
%This stochasticity arises because for a fixed set of macroscopic parameters (e.g., $\kappa, \kappa_*$), different stellar-field realizations produce different micro-image configurations and interference patterns, thus introducing stochasticity in microlensing waveforms.
The many-to-one mapping from stellar-field realizations to the characteristic observables (e.g., $\kappa, \kappa_*$) makes the inference problem inherently stochastic.
Besides the microlensing measurement problem, intrinsic stochasticity may also affect waveforms from post-merger neutron stars and core-collapse supernovae, due to processes such as magnetohydrodynamic (MHD) turbulence, uncertain equations of state, or incomplete theoretical modeling~\citep{2017LRR....20....7P, Baiotti_2017, 2012ARNPS..62..407J, Ott_2009}. Therefore, it is both observationally and theoretically important to identify pathways towards solving the stochastic inference problem. 
%the complexity of the stellar field has so far posed a major obstacle to such investigations.  
In this work, we introduce an artificial intelligence–driven method to identify microlensed Binary Black Holes (BBHs) through their distinctive lensing signatures by employing a set of phenomenological surrogate parameters, such as $l_p$ and $l_{d[M]}$, that capture essential wave-optical features of the waveform.
Here ``surrogate'' means that these parameters are waveform-level descriptors rather than direct macroscopic properties of the stellar field; nevertheless, their correlation with macroscopic quantities such as the total convergence $\kappa$ and stellar microlensing convergence $\kappa_s$ suggests a connection between the inferred waveform features and the underlying stellar field.
This provides a scalable statistical inference solution to the traditionally intractable problem of 
high-dimensional, stochastic microlensing waveforms.

\begin{figure*}[t!]
    \centering
    \includegraphics[width=\linewidth]{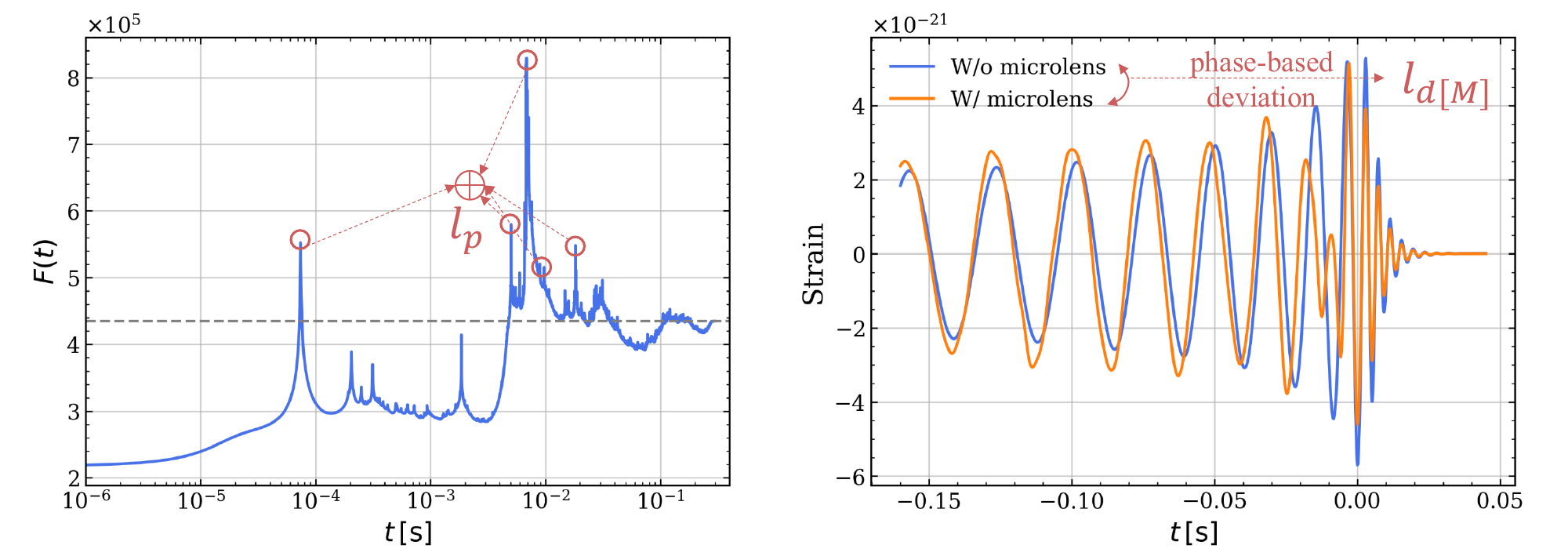}
    \caption{Microlensing surrogate parameters for network inference. The left panel shows the time-domain magnification factor $F(t)$ and the top 10 highest peak values, $l_p$. The right panel shows the time-domain gravitational wave waveform with the microlensing effect (blue) and without the microlensing effect (orange). One can see the deviations ($l_{d[M]}$) caused by microlensing.}
    \label{fig:parameters}
\end{figure*}

\section{Stochastic waveform}
%\pdfbookmark[1]{Non-deterministic waveform}{ndw}
%GWs passing through a stellar field are likely micro-lensed by many nearby stars, so that the final waveform may be written as $h(\theta_{\rm GR}, \theta_{\rm *})$, where $\theta_{\rm GR}$ are the binary parameters and $\theta_{\rm *}$ are the mass and location of all relevant stars. However, since the dimension of $\theta_{\rm *}$ is likely $\ge\mathcal{O}(10^?)$, it is computationally prohibitive to perform parameter estimation (PE) on $\theta_{\rm *}$. A realistic approach is to identify a subset of lensing waveform parameters $\theta_s \in \theta_*$ and perform a join PE with $\theta_{\rm GR}$. This process can be viewed as inference with nondeterministic waveform because $(\theta_{\rm GR}, \theta _s)$ alone cannot uniquely determine the microlensed waveform. Their posterior distribution should be given by
GWs passing through a stellar field are likely to be microlensed by many nearby stars collectively, such that the resulting waveform can be expressed as $h(\theta_{\rm GR}, \theta_{\rm *})$, where $\theta_{\rm GR}$ denotes the standard binary black hole source parameters, including the chirp mass, mass ratio, etc., and $\theta_{\rm *}$ represents the mass and positions of all relevant stars.
However, the estimate $\dim(\theta_*)\gtrsim\mathcal{O}(10^4)$ should be understood as an order-of-magnitude statement rather than a sharp threshold. For the stellar-field configurations considered here, convergence of the diffraction calculation requires retaining at least several thousand relevant microlenses. Since $\theta_*$ contains at minimum one mass and two lens-plane coordinates for each microlens, $N_*\sim$ a few $\times10^3$ already implies $\dim(\theta_*)\sim3N_*\gtrsim\mathcal{O}(10^4)$; the exact number varies with the stellar surface density and the integration region required for convergence. Bayesian inference over the full micro-lensing parameter space $\theta_*$ is computationally prohibitive in practice because of its very high dimension and the intrinsic stochasticity of the waveform.
A more practical approach is to infer a low-dimensional set of surrogate lensing parameters $\theta_s$ that provide a compressed representation of the microlensing effect induced by the full stellar-field configuration $\theta_*$, rather than recovering the stellar masses and positions themselves, and to perform joint parameter estimation over $(\theta_{\rm GR}, \theta_s)$.
This process can be viewed as inference with stochastic waveforms, because for fixed $(\theta_{\rm GR}, \theta_s)$, different stellar-field realizations whose residual freedom is not captured by $\theta_s$ can still produce different waveforms.
The corresponding marginalized posterior distribution is therefore
\begin{align}
P(\theta_{\rm GR}, \theta_s) = \int d\theta_{*\setminus s}\, P(\theta_{\rm GR}, \theta_s, \theta_{*\setminus s})\,.
\end{align}
Here $\theta_{*\setminus s}$ denotes the unresolved micro-lensing degrees of freedom not captured by the surrogate parameters $\theta_s$.
Even for the reduced inference target $(\theta_{\rm GR}, \theta_s)$, conventional Monte-Carlo approaches become impractical, because the waveform still depends stochastically on that unresolved stellar-field freedom, requiring marginalization over many realizations for each likelihood evaluation, which makes conventional MCMC prohibitively expensive.

We therefore adopt simulation-based inference to learn $P(\theta_{\rm GR}, \theta_s)$ directly from simulated data.
Simulation-based inference~\citep{2020PNAS..11730055C} is a class of likelihood-free Bayesian methods that use data generated from a simulator together with neural density estimators to directly learn posterior distributions. This method enables efficient parameter estimation when an analytic likelihood is unavailable for complex physical models.
In this work, we employ neural posterior estimation (NPE)~\citep{dingo2021}, a simulation-based inference technique that learns the posterior distribution from simulated parameter--data pairs using a neural density estimator without requiring an explicit likelihood.
Specifically, we implement NPE using a normalizing flow to compute the joint distribution $P(\theta_{\rm GR}, \theta_s)$.
Note that NPE is not feasible for inference over the full micro-lensing space $\theta_*$ either, because its very high dimension makes it difficult to generate training data that adequately cover the parameter space and thereby enable the normalizing flow to learn the posterior reliably.
Firstly, during training, the detector noise can be randomly simulated using the real detector noise added to each training waveform, i.e., $d=n+h$~\citep{dingo2021}, 
together with the intrinsic stochasticity arising from different stellar-field realizations. The normalizing flow is trained directly on these simulated data without explicitly distinguishing between the two sources of randomness.
%enabling the normalizing flow to handle inference problems involving stochastic detector noise. Since the neural network is agnostic to the physical origin of the noise, only recognizing its presence in the data, it should likewise be able to deal with scenarios where randomness originates from the waveform itself.
Secondly, the normalizing flow is also capable of marginalizing over unconsidered parameters during inference~\citep{vanstraalen2024premergerdetectioncharacterizationinspiraling,PhysRevD.109.123547}. This is because the normalizing flow learns the full joint distribution of all parameters during training. At the inference stage, marginalization over any subset of parameters can be performed efficiently via integration over the learned distribution, without requiring retraining or explicit sampling over the marginalized parameters.

Inspired by \citet{vanstraalen2024premergerdetectioncharacterizationinspiraling,PhysRevD.109.123547}, which perform inference with marginalization over nuisance parameters either by reducing the learned parameter space or by marginalizing at the inference stage, we formulate a proof-of-principle example to examine the behavior of neural posterior estimation under marginalization.
As a proof of principle, we consider the BBH waveform with GR parameter $\theta_{\rm GR}$ and split it into $(\theta_0,\theta_{\rm GR\setminus 0})$.
Here, $\theta_0$ is chosen as the chirp-mass parameter used in the BBH waveform model. The chirp mass, $\mathcal{M}_c=(m_1m_2)^{3/5}(m_1+m_2)^{-1/5}$, is distinct from an individual component mass; source-frame and detector-frame chirp masses are related by $\mathcal{M}_c^{\rm det}=(1+z)\mathcal{M}_c^{\rm src}$. In this proof-of-principle test, $\theta_0$ is introduced only as the parameter to be marginalized over and plays the role of $\theta_{*\setminus s}$ in this example and is not explicitly labeled during NPE training.
The normalizing flow then yields the learned posterior $P_{\rm NPE}(\theta_{\rm GR\setminus 0})$, which can be validated by checking whether it satisfies
\begin{align}
P_{\rm NPE}(\theta_{\rm GR\setminus 0}) \approx P(\theta_{\rm GR\setminus 0}) =\int d \theta_0 P(\theta_{\rm GR})\,.
\end{align}
In this proof-of-principle example, the BBH waveforms are deterministic and microlensing is not included, so $P(\theta_{\rm GR})$ can be obtained with standard Markov chain Monte Carlo (MCMC) analyses and used as a reference for comparison (see Supplementary Material).

Indeed, this setup allows us to validate two complementary implementations of marginalization in neural posterior estimation: (i) marginalization of the full joint posterior obtained from a model trained on the complete parameter space, and (ii) marginalization of the learned posterior $P_{\rm NPE}(\theta_{\rm GR\setminus 0})$. We find that
the posterior distribution $P_{\rm NPE}(\theta_{\rm GR\setminus 0})$ is consistent with both $\int d \theta_0 P(\theta_{\rm GR})$ and $\int d \theta_0 P_{\rm NPE}(\theta_{\rm GR})$, where the training of $P_{\rm NPE}(\theta_{\rm GR})$ corresponds to deterministic waveforms.
Quantitatively, comparing one-dimensional marginalized posteriors for each of the nine parameters, the Jensen--Shannon divergence (JSD) between every pair among $P_{\rm NPE}(\theta_{\rm GR\setminus 0})$, $\int d\theta_0\,P_{\rm NPE}(\theta_{\rm GR})$, and $\int d\theta_0\,P(\theta_{\rm GR})$ remains below $0.1$, with most values around $0.01$ (see Supplementary Material for corner plots and further details).
Although this promising consistency has only been verified in a specific scenario, it suggests that a similar relation may still hold for more general settings, which motivates the application of normalizing flow to infer the posterior distribution of lensing parameters.

\section{Data and training}
%For the micro-lensing signal, 
%\pdfbookmark[1]{Data and training}{dnt}
The microlensed GW data are generated by multiplying the source GW waveform from a BBH with the microlensing magnification in the frequency domain. Unlike in frameworks such as DINGO~\citep{dingo2021}, where the waveform is determined solely by the BBH parameters, these mock microlensed GW data introduce an additional stochastic component. The microlensing signal, which arises from a random distribution of stars or stellar remnants, cannot be uniquely characterized by a few lensing parameters alone, because it still depends on the masses and positions of the many unresolved stars in the field. Consequently, the forward model is intrinsically probabilistic; even for fixed BBH and retained lensing parameters, different random placements of those stars can produce distinct waveforms due to different micro-image configurations, which significantly increases the modeling complexity.

To capture the microlensing signature, we parametrize it by the surrogate set $\theta_s=\{l_p,\,l_{d10},\,l_{d60},\,l_{d120}\}$, as shown in FIG.~\ref{fig:parameters}.
The parameter $l_p$ is calculated as the sum of the 10 highest peaks of the time-domain magnification factor. Since each peak in the magnification corresponds to a distinct micro-image, $l_p$ effectively quantifies the contribution of the 10 most magnified micro-images. The other parameters, $l_{d10}$, $l_{d60}$, and $l_{d120}$, are designed to capture phase-based waveform deviations across low-, intermediate-, and high-mass BBH events (corresponding to binary black hole systems with component masses of $(10\,M_{\odot},\,10\,M_{\odot})$, $(60\,M_{\odot},\,60\,M_{\odot})$, and $(120\,M_{\odot},\,120\,M_{\odot})$). 
Importantly, $l_{d10}$, $l_{d60}$, and $l_{d120}$ are not phases themselves. They are signed, dimensionless mismatch-like statistics constructed from the imaginary part of a normalized noise-weighted inner product between the unlensed waveform and the microlensing-induced waveform difference. We refer to them as ``phase-based'' because this construction is sensitive to the relative phase distortion induced by microlensing: positive and negative values encode opposite signed deviations, while values near zero correspond to negligible phase distortion. This construction also provides a more symmetric and well-behaved feature representation than the conventional always-positive mismatch, which is beneficial for learning-based inference.
%They provide surrogate measures of microlensing-induced phase distortions, with larger values corresponding to stronger deviations between the lensed and unlensed signals in each mass regime.
A detailed definition and discussion of these parameters and distributions are provided in the Supplementary Material~\cite{sup_mat}.

Following the Dingo NPE framework~\citep{dingo2021}, we employ a neural spline flow (NSF)~\citep{nsf_durkan2019neural} as the density estimator, which provides a flexible model for complex, potentially multimodal posteriors.
The NSF is then trained to learn the joint posterior $P(\theta_{\rm GR}, \theta_s)$.
We prepare $5\times 10^5$ microlensed waveforms from an equal number of stellar-field realizations, together with $\sim 5\times 10^6$ vacuum BBH waveforms. Both sets are partitioned into training and held-out evaluation subsets in a 19:1 ratio. During training, each BBH waveform is randomly paired with a microlensing realization. The injection tests reported below and in the Supplementary Material are drawn from the held-out set by rejection sampling so that the accepted events follow an astrophysical population model rather than the uniform training priors.
Microlensing conditions are varied by uniformly sampling $\kappa\in[0.1,0.4]$, $\kappa_s\in[0.1,\kappa/1.2]$, $z_s\in[0.1,2]$, and $z_{\rm L}\in[0.15,z_s]$. The upper limit $z_s\le2$ focuses on the higher-SNR regime where microlensing signatures are more likely to be identifiable, while $\kappa\le0.4$ balances physical coverage against the computational cost of generating large microlensing datasets. Further details and related simulation settings are given in the Supplementary Material.
By exposing the model to these varied conditions, the NSF learns to infer $P(\theta_{\rm GR}, \theta_s)$ from the simulated data, enabling it to recognize microlensing signatures and assess their statistical significance relative to a null (no-microlensing) hypothesis, as quantified below.
Once trained, posterior inference for a new event does not require generating microlensing templates on the fly over the high-dimensional stellar field, which would be the costly step in a direct Monte Carlo analysis of $\theta_*$.
We evaluate the trained network on held-out injection events, as reported in the following section.
%\textcolor{blue}{Still need a paragraph to summarize the training here, as the above paragraphs are about data.}

% 
% The $l_{d[M]}$ is the waveform deviation caused by microlensing.

% is defined as:
% % We define the parameter $l_{d[M]}$ as:
% \begin{equation}
% \label{eq:match}
% l_{d[M]} = \Im \left[ \frac{\langle h_1 \mid h_1 - h_2 \rangle}{\sqrt{\langle h_1 \mid h_1 \rangle \langle h_1 - h_2 \mid h_1 - h_2 \rangle}} \right],
% \end{equation}
% where $\Im$ denotes the imaginary part of the complex quantity. The term $\langle \cdot \mid \cdot \rangle$ represents the noise-weighted inner product. $h_1$ is the unlensed waveform ($h_{\text{U}}$), and $h_2$ is the macro and microlensed waveform ($h_{\text{L}}$).

\section{Parameter estimation results and implications}
In this section, we present the parameter estimation results trained with Neural Posterior Estimation (NPE)~\citep{dingo2021} for microlensed gravitational waves under the specific configurations of third-generation (3G) detector networks. 
While our framework is generally applicable to gravitational-wave observations, we focus on the 3G era because the higher signal-to-noise ratios expected for future detectors make stellar microlensing signatures significantly easier to detect and characterize.
We assume a two-detector network located at the Hanford and Livingston sites, with both employing the Cosmic Explorer (CE)~\citep{reitze2019cosmicexploreruscontribution} design as the Power Spectral Density (PSD).

%\pdfbookmark[1]{Parameter Estimation (PE) results and implications}{pe}

\begin{figure}
    \centering
    \includegraphics[width=\linewidth]{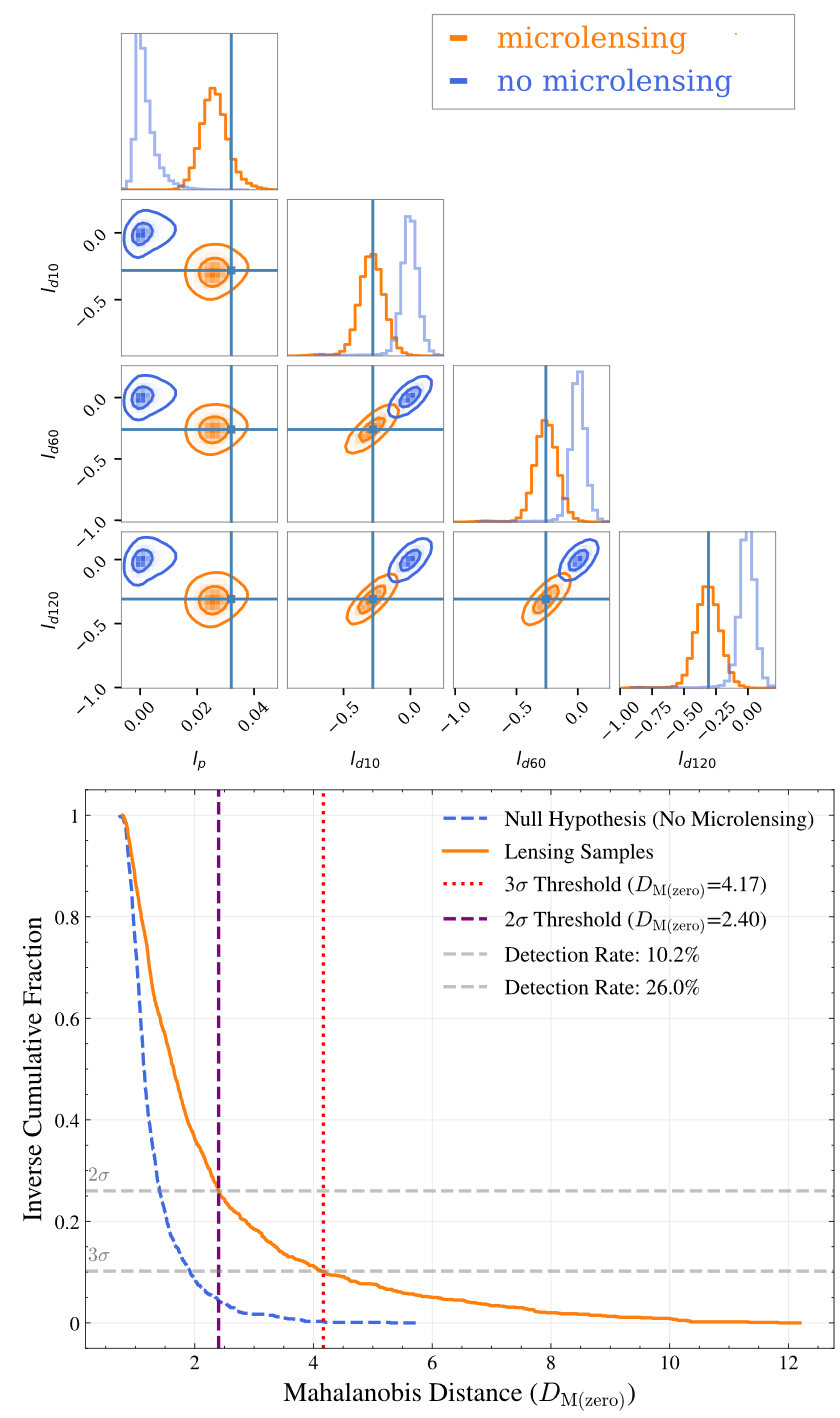}
    \caption{Detection of the microlensing signal. Top: a sampled parameter estimation (PE) result from a microlensing and null-microlensing injection. The $l_p$, $l_{d10}$, $l_{d60}$, and $l_{d120}$ are surrogate parameters, which stand for the peak values in the time-domain magnification factor and microlensing-induced deviations. Bottom: Inverse cumulative fraction of the Mahalanobis distance ($D_{\mathrm{M(zero)}}$) for the lensed samples (orange) and the null hypothesis (dashed blue). The $D_{\mathrm{M(zero)}}$ is calculated relative to the non-lensing origin ($l_p, l_d = 0$). The horizontal dashed line marks the $2\sigma$ and $3\sigma$ threshold derived from the null distribution. At the $2\sigma$ level ($D_{\mathrm{M(zero)}} \ge 2.40$), $26.0\%$ of the lensed population is statistically identifiable, while at the $3\sigma$ level ($D_{\mathrm{M(zero)}} \ge 4.17$), $10.2\%$ can be distinguished from the null hypothesis.} 
    \label{fig:PE_lens_mah_redshift}
\end{figure}

The orange contour in the top panel of FIG.~\ref{fig:PE_lens_mah_redshift} presents the parameter estimation result for an injection containing a distinct microlensing signal. The figure displays the marginalized posterior distributions for the four surrogate lensing parameters, with 50\% and 90\% credible regions.

To evaluate the confidence of the recovered microlensing parameters and the distinguishability of this microlensing event, we designed a null test by preparing comparison events that share the same strong lensing component but without microlensing. This ensures that both injections have nearly the same signal-to-noise ratio (SNR). The blue contour in the top panel of FIG.~\ref{fig:PE_lens_mah_redshift} is the null test result. One can see that the posterior distributions of the microlensing parameters (i.e., $l_p$, $l_{d10}$, $l_{d60}$, and $l_{d120}$) are peaked around zero, consistent with the null microlensing characteristic. In contrast, the microlensed injection (orange) exhibits posterior distributions that are consistently shifted away from zero across all lensing parameters. The one-dimensional posteriors show distinct peaks that are either non-overlapping with ($l_p$) or partially overlapping in the tails of ($l_{d[M]}$) those from the strong lensing only case (blue).

For the two-dimensional posterior distributions, in panels involving the lensing peak parameter $l_p$, such as $(l_p, l_{d10})$, $(l_p, l_{d60})$, and $(l_p, l_{d120})$, the 90\% credible contours of the microlensed (orange) and strong lensing only (blue) cases are clearly separated. While the posteriors involving only deviation parameters, such as $(l_{d10}, l_{d60})$ and $(l_{d60}, l_{d120})$, partially overlap, they still exhibit visibly different centers. This indicates that in this example, $l_p$ provides the most distinctive signal of microlensing, while the deviation parameters contribute additional but subtler discriminative power through their correlated structure.
%Here, it is worth mentioning that the peak parameter $l_p$ can be less than zero. This is due to the definition of $l_p$, which is the peak value in FIG.~\ref{fig:parameters} minus the baseline (black dashed line) value. In some cases, the peak is below the dashed line, making $l_p$ less than zero.

However, not all microlensing waveforms are easily distinguishable, as some microlensing signatures are relatively weak. Therefore, it is essential to assess the distinguishability of microlensed signals quantitatively. 
Therefore, we use Mahalanobis distance~\citep{mclachlan1999mahalanobis, Ezquiaga:2023xfe,PhysRevD.99.043506} to quantify the detectability of an injected signal. Specifically, we calculate the Mahalanobis distance $D_{\mathrm{M(zero)}}$ on the surrogate parameter space relative to the non-lensing origin ($l_p, l_d = 0$) as follows:
\begin{equation}
D_{\mathrm{M(zero)}} = \sqrt{\boldsymbol{\mu}^\top \mathbf{\Sigma}^{-1}\boldsymbol{\mu}},
\end{equation}
where:
\begin{itemize}
    \item $\boldsymbol{\mu}$ is the mean vector of lensing parameters in the posterior distributions,
    \item $\mathbf{\Sigma}$ is the corresponding covariance matrices of lensing parameters in the posterior distributions.
\end{itemize}

By normalizing the displacement of the posterior mean by the inferred covariance, it effectively measures the detectability of microlensing features in the surrogate parameter space. To statistically evaluate the performance of our framework, we apply this metric to an astrophysical population of 1,000 simulated lensed events (see ``Test Dataset'' section in Supplementary Material for more details). The bottom panel of FIG.~\ref{fig:PE_lens_mah_redshift} illustrates the complementary cumulative distribution of $D_{\mathrm{M(zero)}}$ for both the lensed samples and a null-hypothesis (non-lensed) population.
To establish a robust detection criterion, we define significance thresholds based on the null distribution, with $D_{\mathrm{M(zero)}} \ge 2.40$ and $D_{\mathrm{M(zero)}} \ge 4.17$ corresponding to the $95.4\%$ ($2\sigma$) and $99.7\%$ ($3\sigma$) percentiles, respectively. We find that $26.0\%$ and $10.2\%$ of the lensed samples exceed the $2\sigma$ and $3\sigma$ thresholds, respectively (for source redshift $z_s \le 2$).

\begin{figure}
    \centering
    \includegraphics[width=\linewidth]{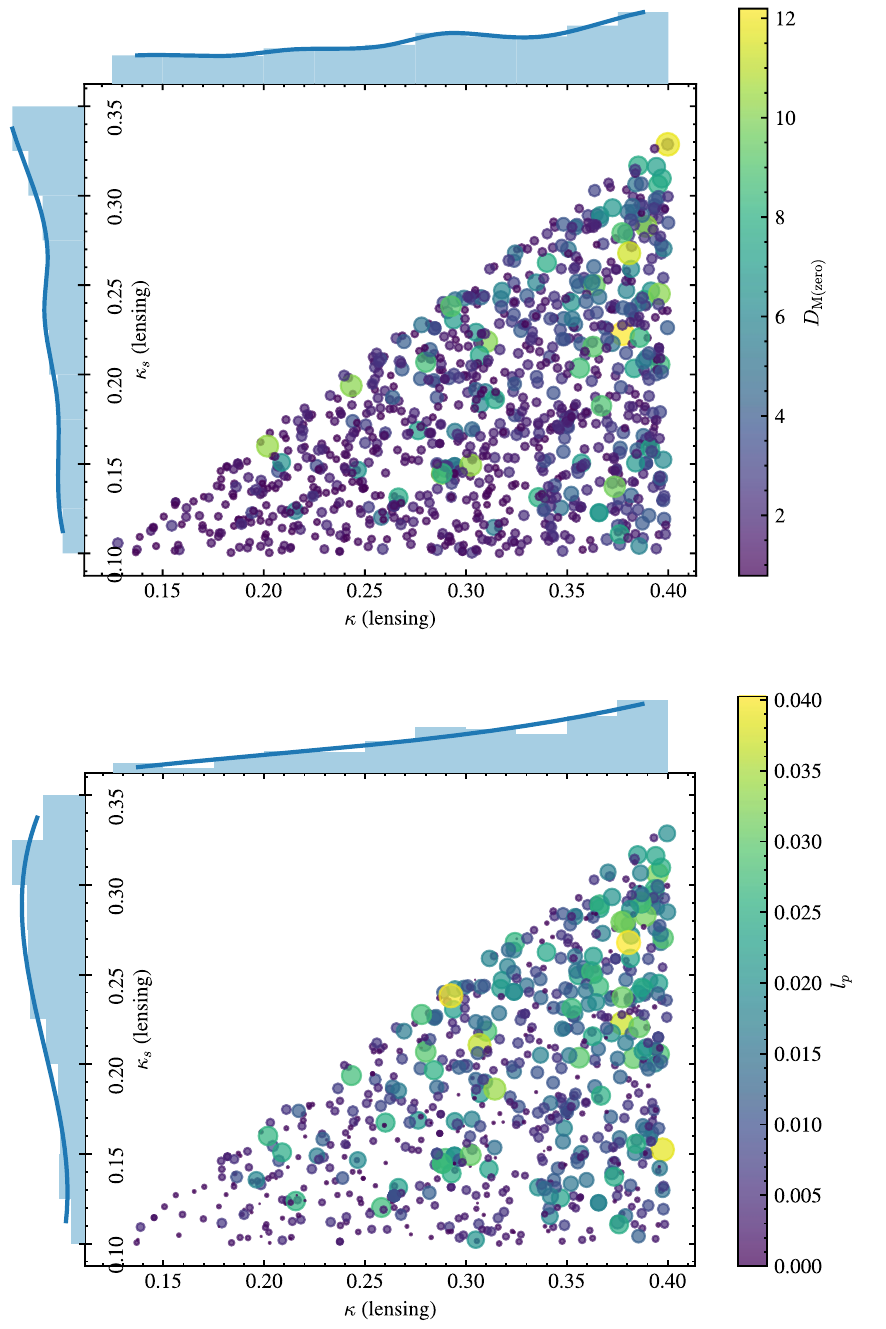}
    \caption{Mahalanobis distance $D_{M(zero)}$ distribution and microlensing surrogate parameter $l_p$ distribution with $\kappa$ and $\kappa_s$.
    Each point represents a specific stellar-field realization; marginal histograms show the projected trends. (Top) Mahalanobis distance $D_{M(zero)}$, where color/size indicate microlensing significance. (Bottom) Inferred surrogate parameter $l_p$, representing peak interference intensity. Both panels exhibit an overall gradient toward the upper-right, reflecting the correlation between wave-optical signatures and macroscopic lens properties ($\kappa, \kappa_s$).
    }
    \label{fig:ka_kas_mah}
\end{figure}

Although the current model does not explicitly predict physical lensing parameters such as the convergence $\kappa$, which reflects the total surface mass density at the SLGW position, and $\kappa_s$, which isolates the portion attributable to compact objects such as stars/remnants, we investigate their correlation with our proposed microlensing discriminators. In particular, we examine how the Mahalanobis distance $D_{\mathrm{M(zero)}}$, which serves as our primary metric for quantifying microlensing detectability, varies with $\kappa$ (x-axis) and $\kappa_s$ (y-axis) (FIG.~\ref{fig:ka_kas_mah}, top panel). The results indicate an overall trend: events with higher values of $\kappa$ and, more prominently, $\kappa_s$ tend to exhibit larger Mahalanobis distances, implying that microlensing signals become more distinguishable as the contribution from the stellar population increases. 
% This provides evidence that the microlensing-sensitive latent representation learned by our network carries predictive power for physical quantities like $\kappa_s$, despite not being directly trained on them.

To further explore this relation, we also visualize the inferred peak lensing parameter $l_p$ across the same $\kappa$–$\kappa_s$ plane (FIG.~\ref{fig:ka_kas_mah}, bottom panel). A similar increasing trend is observed: $l_p$ increases systematically with increasing $\kappa$ and $\kappa_s$, suggesting that the network’s learned representation of lensing structure reflects physical features of the stellar field. These trends indicate that the inferred surrogates are correlated with macroscopic lens properties such as $\kappa$ and $\kappa_s$. We emphasize, however, that the present analysis provides statistical inference only for the surrogate parameters; Figure~\ref{fig:ka_kas_mah} shows a qualitative correlation rather than a quantitative constraint on $\kappa$ or $\kappa_s$. Establishing a mapping that would enable statistical constraints on those macroscopic quantities is left for future work.% A complementary quantitative analysis of this correlation, illustrating the fraction of events with $D_{\mathrm{M}} \ge 3$ across the $\kappa$–$\kappa_s$ parameter space, is presented in Supplementary FIG.~7.

%I recommend to remove this part, because we did not show the figure and I worry it may induce some issues.
%In contrast, other lensing parameters such as $l_{d[M]}$ do not show strong monotonic correlation with $\kappa$ or $\kappa_s$. However, this does not diminish their relevance: while $\kappa$ and $\kappa_s$ describe the lensing surface mass density and the stellar contribution, observables like $l_{d[M]}$ may encode complementary dimensions of microlensing, such as phase-based fluctuations or fine-scale patterns associated with the stellar mass distribution. This highlights the multidimensional nature of microlensing and points toward the benefit of incorporating additional physical descriptors in future modeling.

%Finally, to quantitatively assess the calibration and reliability of the inferred posteriors, we refer to the p–p plot in Supplementary FIG.~6, which provides a consistency check between injected and recovered parameters over multiple realizations.

\section{Discussion}
%\pdfbookmark[1]{Discussion}{dis}
%~The successful identification of %strongly lensed gravitational waves (SLGWs) and their associated 
%microlensing signatures is a critical prerequisite for exploiting their potential in cosmological and astrophysical studies. In recent years, several approaches have been proposed for strongly lensed gravitational wave identification~\citep{Haris:2018vmn, Barsode2024FastWaves, Kim:2020xkm, Goyal:2021hxv, Lo:2021nae, Liu:2020par, Janquart:2021qov, Janquart_2023, Dai:2017huk, Wang:2021kzt, Janquart:2021nus, Ezquiaga:2023xfe, Wright:2025nod, Chakraborty:2024net}, Monte Carlo-type methods to identify isolated microlensing events~\citep{Wright:2021cbn, Nerin_2025, chakraborty2025modelindependentchromaticmicrolensingsearch, PhysRevD.111.084019, 2025arXiv250520996Q}, and template-free methods to recover microlensing-induced distortions~\citep{Dai:2018enj, Zumalacarregui:2024ocb, Shan:2023ngi, Chakraborty:2024mbr}.
%However, to date, no study has performed parameter estimation for SLGWs affected by a stellar microlensing field. 
As discussed in the Introduction, although some studies have developed methods for strongly lensed gravitational-wave identification and for microlensing detection or reconstruction in simpler lens models, no study has addressed parameter estimation for stochastic microlensing produced by large stellar fields, where the waveform depends on thousands or more relevant stars.
A major challenge is the inherently stochastic nature of the lensed waveform, even for fixed values of the strong-lensing convergence $\kappa$, shear $\gamma$, microlensing surface density $\kappa_s$, and the microlens mass function. In this work, we address this challenge by employing a normalizing flow, 
offering a novel framework to tackle the stochasticity of microlensed waveforms.
In this proof-of-principle study, microlensing detection is implemented by inferring the surrogate parameters for each event and applying the Mahalanobis distance statistic $D_{\mathrm{M(zero)}}$, calibrated against a null (no-microlensing) population, to select microlensing-identifiable systems from a simulated lensed sample. Because the signatures considered here require the higher signal-to-noise ratios expected for third-generation detectors, this selection procedure is demonstrated on simulated injections rather than applied to present observational data.
Here, ``SLGW'' is used only to describe the strong-lensing configuration or source population. The statistic developed in this work identifies microlensing signatures in an individual macroimage and does not identify strongly lensed multiple-image pairs.

Our results indicate that a normalizing flow can recover the microlensing surrogate parameters and distinguish the microlensing effect from a null-microlensing background in the regime of sufficiently strong microlensing signatures. 
According to our results shown in FIG.~\ref{fig:PE_lens_mah_redshift},
based on the predicted detection rate of $\sim40$–$1000$ SLGWs per year in the 3G~\citep{2014ASSL..404..333P,reitze2019cosmicexploreruscontribution} detector era~\citep{Li:2018prc, Xu:2021bfn}, 
considering that our analysis is restricted to $z_s\leq2$, which accounts for approximately half of the strongly lensed GW population in the adopted population model~\citep{Xu:2021bfn}, the above rate corresponds to roughly $2$–$51$ events annually with a $3\sigma$ detection. At the less conservative $2\sigma$ threshold, this corresponds to roughly $5$–$130$ identifiable events annually.
%this corresponds to roughly $4$–$102$ events annually with a $3\sigma$ detection. 
This estimate is dependent on the strong-lensing galaxy population and the BBH population model. Furthermore, although type-II (saddle-point) macroimages in the standard Morse classification, in contrast to type-I minimum images~\citep{Dai:2017huk}, were not included in this analysis, they typically exhibit stronger and more easily identifiable microlensing effects, suggesting a potentially larger fraction of detectable cases. They are therefore mentioned only as a possible extension and are not part of the detection fractions reported here.

%Note that the surrogate parameters are designed to better describe the microlensing signatures in the waveform. They are not commonly used ``lensing variables'', such as $\kappa, \kappa_s$, but they are correlated with them. This correlation is evident in FIG.~\ref{fig:ka_kas_mah}, which shows that the bin-averaged peak value are positively correlated with both the strong lens convergence ($\kappa$) and in particular, the microlensing density ($\kappa_s$) which is also a measure of the density of the stellar field. This finding is physically intuitive, as the statistical significance of the microlensing effect is expected to increase with higher values of $\kappa$ and $\kappa_s$. This correlation also suggests the possibility of constraining these physical parameters through a hierarchical Bayesian inference framework, or a dedicated mapping between waveform surrogate variables and more detailed mass distribution in the microlensing field. There are likely to be more effective sets of parameters to separately characterize the microlensing waveform modulation and the lensing field. Constructing an informative mapping between them will be an interesting direction to pursue for future studies.

%In this work, we use surrogate parameters to better describe the microlensing signatures in the waveform rather than the commonly used ``lensing variables'', such as $\kappa$ and $\kappa_s$. This choice is motivated by the capabilities of the neural network, as we found that the common ``lensing variables'' are difficult for the current normalizing flow configuration to learn.
We have adopted waveform-level surrogate parameters, rather than attempting to infer physical lensing parameters such as $\kappa$ and $\kappa_s$ directly, because the present analysis targets the recoverable microlensing signature in the waveform. Inferring the macroscopic lens quantities is left for a subsequent mapping or hierarchical analysis.
At the population level, interesting correlations exist between the physical lensing parameters $\kappa,\kappa_s$ and the surrogate parameters. This correlation is evident in FIG.~\ref{fig:ka_kas_mah}, which shows that the bin-averaged peak value is positively correlated with both the strong-lensing convergence ($\kappa$) and, in particular, the microlensing density ($\kappa_s$), which is also a measure of the stellar field density. This behavior is physically intuitive, as the statistical significance of the microlensing effect is expected to increase with higher values of $\kappa$ and $\kappa_s$. 
These correlations indicate that a physically informed mapping between surrogate parameters and lensing quantities may enable a more optimal characterization of microlensing waveform modulation and the stellar field.

Our approach can be extended naturally to more general stochastic waveform inference problems in gravitational wave astronomy, such as post-merger neutron stars, core-collapse supernovae, and extreme mass-ratio inspirals in turbulent Active Galactic Nuclei (``Wet EMRIs" \cite{Pan:2021ksp,Pan:2021oob,Sun:2025lbr}). For core-collapse supernovae and neutron-star post-merger signals in particular, SBI can be useful precisely because it does not require an explicit or tractable likelihood, or a simple deterministic waveform model. Instead, it can learn directly from ensembles generated by stochastic numerical simulators and marginalize over latent physical variability. Uncertainties associated with the microphysics, equation of state, turbulence, and, for core-collapse supernovae, even different explosion outcomes can in principle be incorporated into the simulation ensemble. The essential requirement is that the simulations span the relevant physical variability sufficiently well; SBI cannot compensate for important physics that is entirely absent from the forward simulations. Major challenges remain before a straightforward application to all these problems, including the need to generate sufficiently large training datasets
and to develop appropriate stochastic waveform representations for different physical scenarios. For the microlensing application considered here, a more complete characterization of the detectable population will require extending the analysis to higher-redshift sources, where the reduced signal-to-noise ratios may make microlensing signatures more difficult to identify.
%and the extension of the source population to higher redshifts beyond our current $z_s \le 2$ cut. 
Furthermore, applying this framework to real observations will require careful assessment of potential out-of-distribution effects, where features absent from the training set, such as waveform modeling uncertainties, calibration systematics, or non-astrophysical noises, could be partially captured by the surrogate parameters and mimic microlensing signatures.
%Furthermore, it is important to address the concern that our surrogate parameters might ``absorb'' unmodeled physical effects or waveform systematic errors, leading to potential false alarms. 
Unlike conventional parameter estimation, where parameters often have direct physical interpretations, surrogate parameters are designed to capture complex stochastic effects and therefore require careful validation against realistic data. Addressing this challenge will require future studies incorporating more realistic noise models, waveform uncertainties, and out-of-distribution tests.
%This is an inherent challenge in gravitational-wave astronomy, as seen in complex events like GW231123 where waveform uncertainties allow for multiple interpretations. 
As the first step in this series of work, we have demonstrated that the stochastic waveform problem can be effectively addressed via NPE. In future studies, we plan to use machine learning techniques to identify even more robust feature representations beyond our current hand-picked parameters, which we expect will further enhance the precision and scalability of SBI-based stochastic modeling.
The present analysis already points to this as one of the most promising avenues for exploration.

    %==========================================================================
    \begin{acknowledgments}
        H.Y. is supported by the National Natural Science Foundation of China (Grant 12573048).
        Z.S. is supported by the National Natural Science Foundation of China (Grant 62402274), Start-up Funding of Fuzhou University (Grant XRC-25164), and the Postdoctoral Fellowship Program of China Postdoctoral Science Foundation (Grant GZC20231304). % Zhaoqi Su
        X.S. acknowledges support from Shuimu Tsinghua Scholar Program (No. 2024SM199) and the China Postdoctoral Science Foundation (Certificate Number: 2025M773189).
        X.S. and S.M. acknowledge support from the National Science Foundation of China (Grant No. 12133005).
        Y.L. is supported by the National Natural Science Foundation of China (Grant 62125107). % Yebin Liu

        % \red{This material is based upon work supported by NSF's LIGO Laboratory which is a major facility fully funded by the National Science Foundation.}
    \end{acknowledgments}

\bibliographystyle{apsrev4-2}
\bibliography{references}

\begin{center}
  \textbf{\large Supplemental Materials}
\end{center}
\setcounter{equation}{0}
\setcounter{figure}{0}
\setcounter{table}{0}
\setcounter{section}{0}
\setcounter{page}{1}
\makeatletter
\renewcommand{\theequation}{S\arabic{equation}}
\renewcommand{\thefigure}{S\arabic{figure}}
\renewcommand{\thetable}{S\arabic{table}}
\renewcommand{\thesection}{S\arabic{section}}
\makeatother

\begin{figure*}
    \centering
    \includegraphics[width=\linewidth]{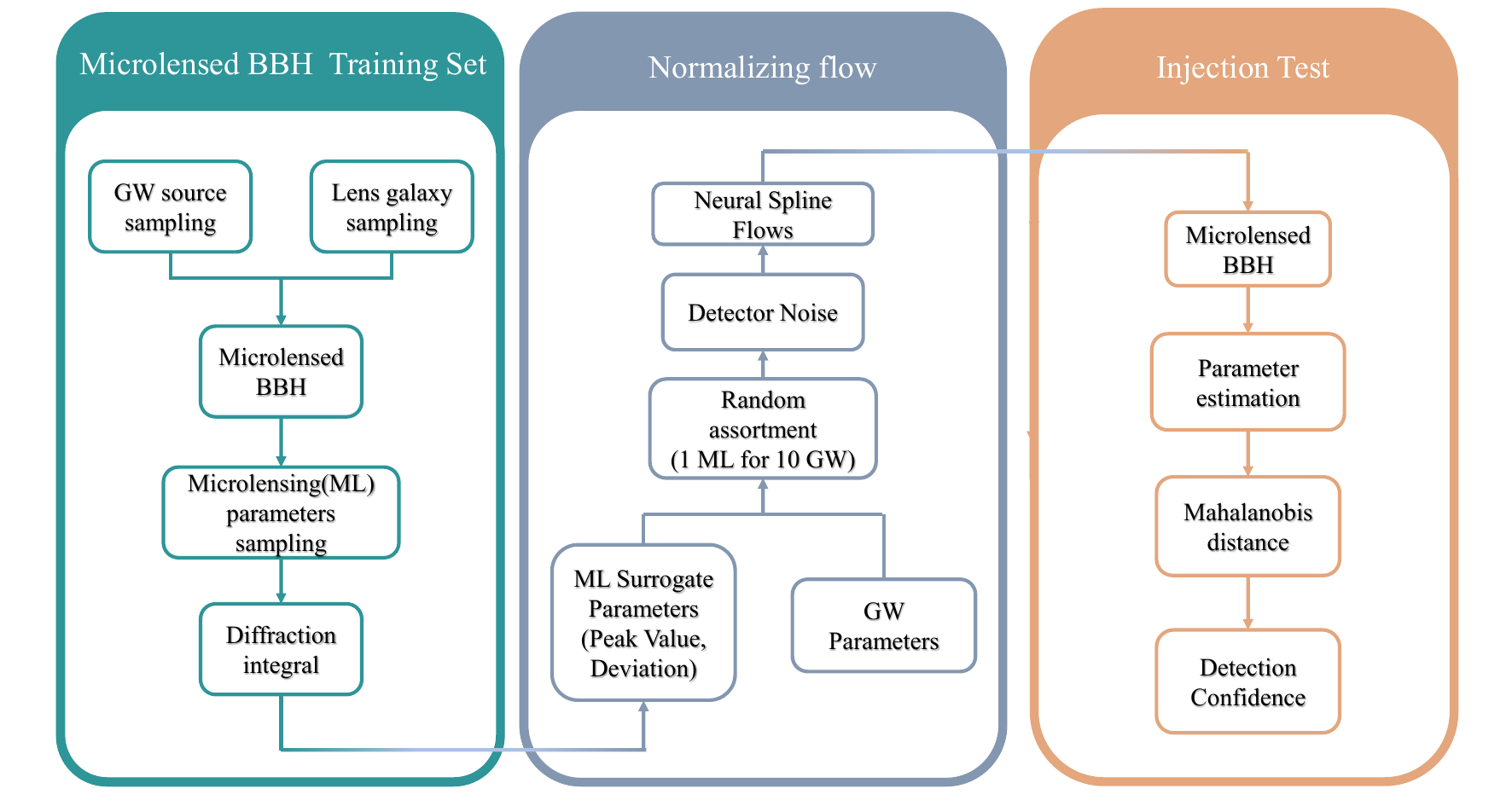}
    \caption{Flowchart of our pipeline. The pipeline consists of three modules. In the microlensed training set, source gravitational-wave signals are generated using \texttt{LALSimulation}~\cite{LALSimulation_python}, while microlensing signals are produced with the SLBBH Mock Data generation pipeline. In the normalizing flow module, we train a Neural Spline Flow~\citep{nsf_durkan2019neural} to learn the joint posterior distribution of both BBH and microlensing parameters. Finally, in the injection test, microlensed signals are injected and the significance of microlensing detection is quantified using the Mahalanobis distance.}
    \label{fig:flowchart}
\end{figure*}

\begin{figure*}
    \centering
    \includegraphics[width=\linewidth]{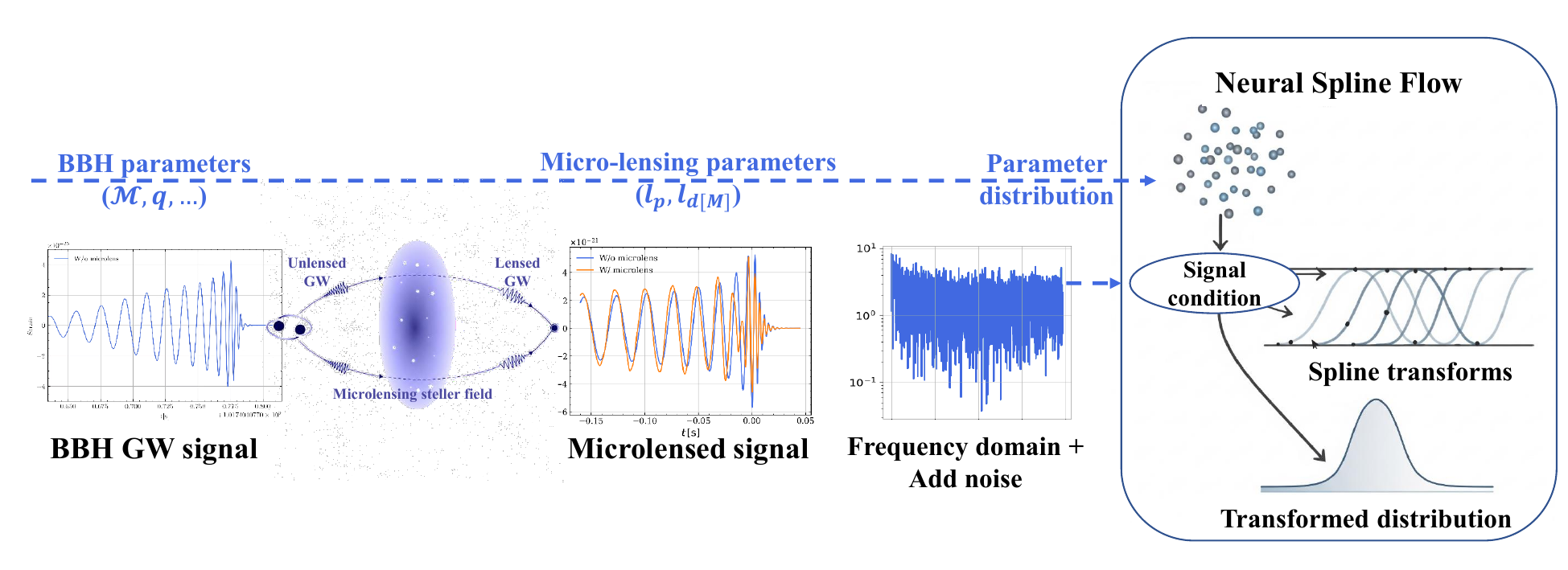}
    \caption{Microlensed BBHs and the detection method with neural networks.}
    \label{fig:pipeline}
\end{figure*}

%
%1. parameters pdf
%2. 

\section{Flowchart}

%We demonstrate the flowchart of our pipeline, as shown in Fig.~\ref{fig:flowchart}. For the training set, the BBH signals generated using \texttt{LALSimulation}~\cite{LALSimulation_python} and the microlensing signal generated using our SLBBH Mock Data generation pipeline (illustrated in the Method Section) are pre-generated. The peak value $l_p$ and the deviation value $l_{d[M]}$ are also pre-calculated in the dataset generation pipeline. By uniformly sampling over $\kappa$ and $\kappa_s$ in the SLBBH Mock Data generation pipeline, the training set provides diverse microlensing conditions, allowing the network to learn waveform variations more effectively. 
We illustrate the overall workflow of the pipeline in FIG.~\ref{fig:flowchart} and FIG.~\ref{fig:pipeline}. The training set consists of BBH signals generated using \texttt{LALSimulation}~\citep{LALSimulation_python} and microlensed signals produced with our microlensed binary black hole Mock Data generation pipeline, described in the following section. The lensing parameters, including the peak value $l_p$ and deviation values $l_{d[M]}$, are pre-computed during data generation.

%During training, due to the multi-dimensional characteristic of the microlensing signals, we first use the BBH signals for SVD basis extraction, and project the whole microlensed waveform onto the SVD space for dimension reduction. For each training epoch, the BBH signal will be shuffled and used once each, while the microlensing signals are randomly paired with the BBH signal under the redshift constraint. On average, each microlensing signal will be paired 10 times in one epoch. With the noise generated from the CE spectrum density (removed glitch), the signals are then fed into the neural spline flow~\cite{nsf_durkan2019neural} for both BBH parameter and microlensing parameter estimation. After training, as mentioned in the Results Section, we use a modified Mahalanobis distance for detecting the confidence of whether the microlensing signal exists.
During training, we first extract a reduced basis from the BBH signals using singular value decomposition, which captures the dominant modes of waveform variation in the absence of microlensing. The full set of microlensed waveforms is then projected onto this basis to reduce dimensionality while retaining relevant physical features. The training inputs are constructed by adding simulated noise drawn from the CE power spectral density~\citep{reitze2019cosmicexploreruscontribution} (with glitches removed). These inputs are passed to the Neural Spline Flows~\cite{nsf_durkan2019neural}, which jointly estimates both the BBH source parameters and the microlensing parameters. After training, the Mahalanobis distance described in the main paper ``Parameter Estimation results and implications'' section is used to quantify the statistical significance of microlensing detection.

% \section{Method}
% \label{subsec:method}
% %\textcolor{blue}{explain the architecture.}

\section{Data Preparation}

\begin{figure*}
    \centering
    \includegraphics[width=\linewidth]{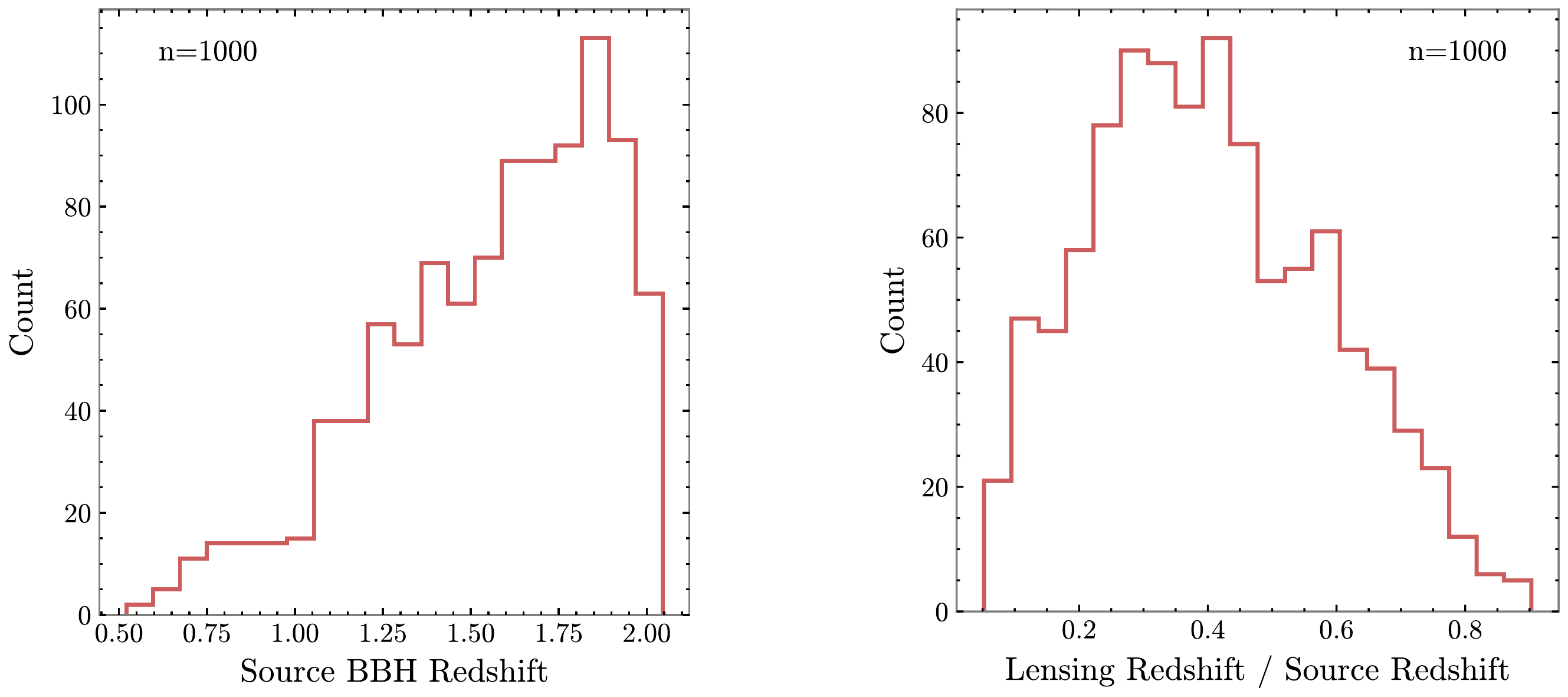}
    \caption{Astrophysical distribution of the sampled 1000 injections. The left panel shows the source redshift histogram distribution, and the right panel shows the ratio of the lens redshift to the source redshift distribution.}
    \label{fig:astrophysics}
\end{figure*}

\subsection{Training Dataset}
\label{subsub:train_set}
This subsection details the training-set generation block of the pipeline in FIG.~\ref{fig:flowchart}.
Our training dataset was generated by uniformly sampling the following parameters: strong lensing convergence ($\kappa$), stellar microlensing convergence ($\kappa_s$), source redshift ($z_s$), and lens redshift ($z_\mathrm{L}$).

The parameter ranges were defined as follows:
%\begin{itemize}
%    \item $\kappa \propto \mathrm{U} (0.1, 0.4)$
%    \item $\kappa_s \propto \mathrm{U} (0.1, \kappa/1.2)$
%    \item $z_s \propto \mathrm{U} (0.1, 2.0)$
%    \item $z_\mathrm{L} \propto \mathrm{U} (0.15, z_s)$
%\end{itemize}

\begin{table}[H]
\caption{Lensing and redshift parameter ranges for the training dataset.}
\label{tab:train_lensing_params}
\begin{ruledtabular}
\begin{tabular}{lcc}
Parameter & Range & Distribution \\
$\kappa$ & $[0.1,\,0.4]$ & $\mathrm{U}(0.1,\,0.4)$ \\
$\kappa_s$ & $[0.1,\,\kappa/1.2]$ & $\mathrm{U}(0.1,\,\kappa/1.2)$ \\
$z_s$ & $[0.1,\,2.0]$ & $\mathrm{U}(0.1,\,2.0)$ \\
$z_\mathrm{L}$ & $[0.15,\,z_s]$ & $\mathrm{U}(0.15,\,z_s)$ \\
\end{tabular}
\end{ruledtabular}
\end{table}

Here, the upper limit of $z_s\leq2$ is adopted to focus on the regime where microlensing signatures are more likely to be identifiable due to higher signal-to-noise ratios. The upper limit of $\kappa\leq0.4$ is chosen to balance the physical parameter range and computational cost, as larger convergence values substantially increase the complexity of microlensing simulations and the required data-generation time.

%Here $\mathrm{U}$ denotes uniform distribution. 
The strong lensing convergence, denoted by $\kappa$, describes the dimensionless surface mass density at the location of a strong lensing image. It is defined as the ratio of the surface mass density, $\Sigma$, to the critical surface density, $\Sigma_\text{crit}$:
\begin{equation}
    \kappa = \frac{\Sigma}{\Sigma_\text{crit}}.
\end{equation}

The convergence of the microlensing stellar field, denoted by $\kappa_s$, is defined as:
\begin{equation}
    \kappa_s = \frac{N\pi}{S},
\end{equation}
where $N$ is the number of microlenses. The lens-plane and source-plane coordinates are expressed in units of the Einstein radius $R_{\rm E}$ corresponding to the average microlens mass $\mathrm{M}_{\rm L}$. Consequently, the region area $S$ is measured in units of $R_{\rm E}^2$; equivalently, after this normalization, the numerical coordinate values and the numerical value of $S$ are dimensionless. This is the convention used in the definition of $\kappa_s=N\pi/S$.

The upper limit for $\kappa_s$ ensures that the total convergence from discrete objects ($1.2\kappa_s$, assuming a 20\% remnant fraction) does not exceed the total convergence $\kappa$.

Additionally, 
we assume that the strong lensing shear amplitude satisfies $|\gamma|=\kappa$, 
%we assume that the strong lensing shear equals the convergence ($\gamma = \kappa$), 
which is consistent with a singular isothermal elliptical (SIE) lens model~\citep{1994A&A...284..285K} 
lens model. Although this simplification neglects external shear and more complex lens structures, it serves as a suitable approximation for this proof-of-principle study and does not impact our primary conclusions.

For the BBH parameters, we randomly assign values for the component masses ($m_1$, $m_2$), inclination angle ($\iota$), polarization angle ($\psi$), right ascension ($\alpha$), declination ($\delta$), merger time ($t_c$), and dimensionless spins ($a_1$, $a_2$) according to the following probability distributions:
%\begin{itemize}
%    \item The mass ratio $q$ is sampled from a uniform distribution $\mathrm{U}(0.125, 1)$, and the chirp mass $\mathcal{M}_c$ from $\mathrm{U}(15, 150)\,\mathrm{M}_\odot$, with the component masses $m_1$ and $m_2$ restricted to the range $[10, 120]\,\mathrm{M}_\odot$. These ranges follow the BBH waveform simulation settings adopted in the Dingo NPE framework~\citep{dingo2021} and its publicly available implementation.
%    \item The inclination angle follows the distribution $p(\iota)\propto \sin(\iota)$, with $\iota \in [0, \pi]$.
%    \item The polarization angle is uniformly distributed: $p(\psi)\propto \mathrm{U}(0,\pi)$.
%    \item The right ascension is uniformly distributed: $p(\alpha)\propto \mathrm{U}(0,2\pi)$.
%    \item The declination follows the distribution $p(\delta)\propto \cos(\delta)$, with $\delta \in [-\pi/2, \pi/2]$.
%    \item The merger time is uniformly distributed: $p(t_c)\propto \mathrm{U}(t_\mathrm{min}, t_\mathrm{max})$, where the simulation spans 1 years.
%    \item We set the dimensionless spin magnitudes $\chi_1$ and $\chi_2$ as 0.
%    %\item The dimensionless spin magnitudes are uniformly distributed between 0 and 0.99: $p(a_1)\propto \mathrm{U}(0, 0.99)$.
%    %\item Similarly, $p(a_2)\propto \mathrm{U}(0, 0.99)$.
%\end{itemize}

\begin{table}[H]
\caption{BBH parameter ranges and priors for the training dataset.}
\label{tab:train_bbh_params}
\begin{ruledtabular}
\begin{tabular}{lcc}
Parameter & Range & Distribution \\
$q$ & $[0.125,\,1]$ & $\mathrm{U}(0.125,\,1)$ \\
$\mathcal{M}_c$ & $[15,\,150]\,\mathrm{M}_\odot$ & $\mathrm{U}(15,\,150)\,\mathrm{M}_\odot$ \\
$m_1$, $m_2$ & $[10,\,120]\,\mathrm{M}_\odot$ & from $q$, $\mathcal{M}_c$; restricted to range \\
$\iota$ & $[0,\,\pi]$ & $p(\iota)\propto\sin\iota$ \\
$\psi$ & $[0,\,\pi]$ & $\mathrm{U}(0,\,\pi)$ \\
$\alpha$ & $[0,\,2\pi]$ & $\mathrm{U}(0,\,2\pi)$ \\
$\delta$ & $[-\pi/2,\,\pi/2]$ & $p(\delta)\propto\cos\delta$ \\
$t_c$ & 1 yr span & $\mathrm{U}(t_\mathrm{min},\,t_\mathrm{max})$ \\
$\chi_1$, $\chi_2$ & $0$ & fixed \\
\end{tabular}
\end{ruledtabular}
\end{table}
These BBH priors follow the waveform-generation settings adopted in the Dingo NPE framework~\citep{dingo2021} and its publicly available implementation, including the mass priors ($\mathcal{M}_c\in[15,150]\,\mathrm{M}_\odot$ and $m_1$, $m_2$ restricted to $[10,120]\,\mathrm{M}_\odot$), etc.
And then, we use \texttt{LALSimulation}~\citep{LALSimulation_python}, employing the IMRPhenomXPHM waveform model~\citep{Pratten_2021}, to generate the corresponding GW signals.

For the simulation of the microlensing field, we follow the procedures outlined in Refs.~\citep{Chen:2021ftm, Zheng:2022vfq,Shan:2022xfx}. 
We employ the Salpeter initial mass function~\citep{1955ApJ...121..161S} to model the stellar mass distribution, where the range is set to [0.1, 1.5] solar masses, consistent with the value used by Diego \textit{et al.}\citep{Diego:2021mhf}. This interval specifies the stellar component of our adopted microlens population and is not intended as a hard cutoff on the full compact-object population. Compact remnants are modeled separately through the initial-final mass relation described below, extending the microlens population beyond the $1.5\,\mathrm{M}_\odot$ upper bound of the stellar component. In addition to stars, we account for remnant objects within the microlensing field. We adopt the initial-final mass relation from Spera \textit{et al.}\citep{2015MNRAS.451.4086S} (ranging from 0.1 solar masses to 28 solar masses) and assume that remnant objects contribute 20\% to the total stellar mass density~\citep{Meena:2022unp}.

To calculate the frequency-dependent magnification caused by microlensing, we utilize the algorithm presented in~\citet{Shan:2022xfx}, which involves evaluating the Fresnel-Kirchhoff diffraction integral~\citep{1992grlebookS}:

\begin{equation}
\label{eq:DiffInter}
F(\omega, \boldsymbol{y})=\frac{2 G \mathrm{M}_\mathrm{L}\left(1+z_\mathrm{L}\right)} \omega{\pi c^{3} i} \int_{-\infty}^{\infty} d^{2} x \exp \left[i \omega t(\boldsymbol{x}, \boldsymbol{y})\right]\;,
\end{equation}
where $F(\omega, \boldsymbol{y})$ represents the wave optics magnification factor, $\omega$ is the gravitational wave's circular frequency, and we use the frequency domain with [20, 1024] Hz. 
Note that we restrict the analysis to 20–1024 Hz for computational efficiency. This choice is independent of our focus on the 3G era, which is motivated primarily by the higher signal-to-noise ratios expected for future detectors rather than their extended low-frequency sensitivity.
$\boldsymbol{y}$ is its position in the source plane (normalized by the Einstein radius). $\mathrm{M}_\mathrm{L}$ and $z_\mathrm{L}$ represent the average microlens mass and the redshift of the strong lensing galaxy, respectively, $\boldsymbol{x}$ denotes the lens plane coordinates, and $t(\boldsymbol{x}, \boldsymbol{y})$ is the time-delay function, defined as:

\begin{equation}
\begin{split}
\label{eq:TimeDelay}
t(\boldsymbol{x},\boldsymbol{x}^{i},\boldsymbol{y}=0)&=\underbrace{\frac{k}{2}\left((1-\kappa+\gamma) x_{1}^{2}+(1-\kappa-\gamma) x_{2}^{2}\right)}_{t_\text{s}(\kappa,\gamma,\boldsymbol{x})} \\
&-  \underbrace{\left[\frac{k}{2}\sum_{i}^{N} \ln \left(\boldsymbol{x}^{i}-\boldsymbol{x}\right)^{2} + k\phi_{-}(\boldsymbol{x})\right]}_{t_\text{m}(\boldsymbol{x},\boldsymbol{x}^{i})}
\end{split}
\end{equation}
Here, $k=4 G \mathrm{M_{L}}(1+z_\mathrm{L})/c^3$ and $\boldsymbol{x^{i}}$ represents the coordinate of the $i$th microlens.

We define the macro image position as the coordinate origin ($\boldsymbol{y} = 0$). $\phi_{-}(\boldsymbol{x})$ represents the contribution from a negative mass sheet, introduced to compensate for the mass contribution from the microlenses and maintain a constant total convergence, $\kappa$~\citep{Wambsganss1990, Chen:2021ftm, Zheng:2022vfq}. The terms $t_\text{s}(\kappa, \gamma, \boldsymbol{x})$ and $t_\text{m}(\boldsymbol{x},\boldsymbol{x}^{i})$ represent the time delays due to the macro lens and the microlenses, respectively. With the completion of these steps, we have successfully generated all necessary components for our training data set.

%\begin{figure*}[t!]
%    \centering
%    \includegraphics[width=\linewidth]{fig/parameters.pdf}
%    \caption{Microlensing surrogate parameters for network inference. The left panel shows the time-domain magnification factor (defined in Eq.~(\ref{eq:TimeDomainMag})) and the top 10 highest peak values, $l_p$. The right panel shows the time-domain gravitational wave waveform with the microlensing effect (blue) and without the microlensing effect (yellow). One can see the deviations ($l_{d[M]}$) caused by microlensing.}
%    \label{fig:parameters}
%\end{figure*}

\subsection{Test Dataset}

\begin{figure*}
    \centering
    \includegraphics[width=\linewidth]{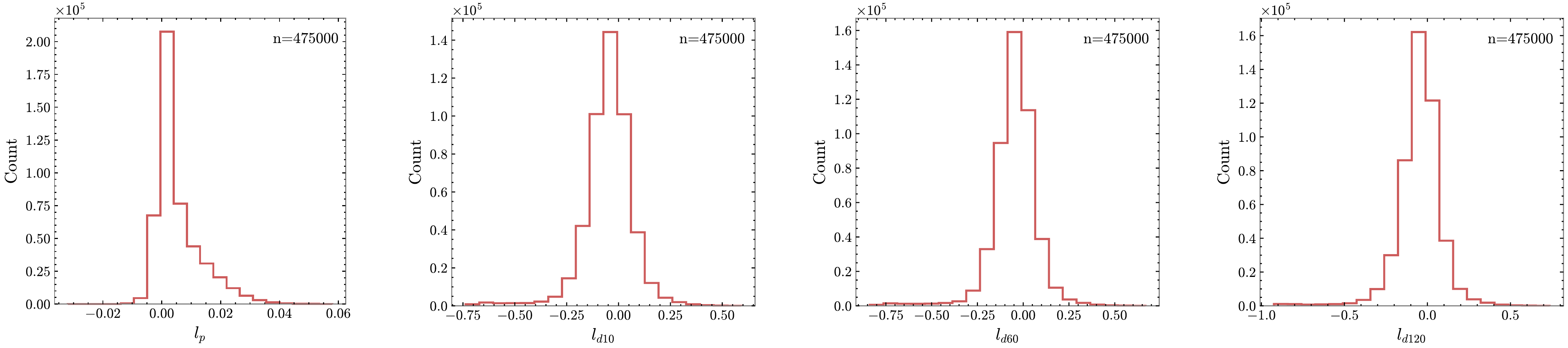}
    \caption{Lensing parameter distribution in the training dataset. The $l_p$, $l_{d10}$, $l_{d60}$, and $l_{d120}$ are surrogate parameters, which stand for the peak values in the time-domain magnification factor and microlensing-induced deviations.}
    \label{fig:lp_ld_distrib}
\end{figure*}

This subsection details the injection-test data used in the final block of FIG.~\ref{fig:flowchart}.
To assess the performance of our methodology, we construct a test dataset of 1000 injection events.
These events are drawn from the held-out evaluation subset of the pre-generated catalogs by rejection sampling, so that the accepted sample follows the astrophysical model of~\citep{Haris:2018vmn,Xu:2021bfn} rather than the uniform training priors.
The selection process unfolds as follows:

\begin{itemize}
  \item We generate BBH redshifts ($z_s$) by sampling from a theoretical BBH merger rate model. This model assumes that the merger rate is proportional to the star formation rate (SFR) with a time delay of $\Delta t = 50 \mathrm{Myr}$ between star formation and subsequent BBH formation. Refer to Appendix B of~\citet{Xu:2021bfn} for a detailed exposition of this model.
  \item For each simulated BBH event, we follow the sampling distributions described in the ``Training Dataset'' section, except for the component masses, which are drawn from a power-law distribution with a peak, $(m_1, m_2) \sim \mathrm{power\ law + peak}$~\citep{LIGOScientific:2018jsj}.
  \item For each BBH event at redshift $z_s$, we compute the multiple-image optical depth, $\tau(z_s)$, using the singular isothermal sphere (SIS) model, as described in ~\citet{Haris:2018vmn}. We then draw a random number from a uniform distribution between 0 and 1. If the calculated optical depth exceeds the random number, we classify the event as an SLGW event; otherwise, it is excluded.
  \item For the selected SLGW events, we sample $\kappa$ and $\kappa_s$ uniformly within the parameter space.
  %\item {\red{ka kas not astro distrib }}For the selected SLGW events, we model the lensing galaxy as a SIE model~\citep{1994A&A...284..285K} and utilize the \texttt{Lenstronomy} software package~\citep{2018PDU....22..189B, 2021JOSS....6.3283B} to solve the lens equation. The velocity dispersion ($\sigma_v$) and axis ratio ($q$) of the SIE are sampled from the observed distribution of the Sloan Digital Sky Survey (SDSS) galaxy population~\citep{2015ApJ...811...20C}. We note that~\citet{2015ApJ...811...20C} contains a typographical error in the axis ratio parameter, and we employ the corrected form as presented in~\citet{Wierda:2021upe}. The sampling details for these parameters, as well as the lens redshift and source-plane location, are detailed in Appendix A of~\citet{Haris:2018vmn}.
\end{itemize}

We then use a detector network containing two CE detectors, situated in Livingston and Hanford.

Finally, for the microlensing stellar field and BBH source parameters, we use the same method described in the ``Training Dataset'' section. Following the above procedure, we obtain 1000 accepted events for the injection tests. We show the sampled source BBH redshift distribution (FIG.~\ref{fig:astrophysics}, left panel) and the proportion of lensing and source redshift distribution (FIG.~\ref{fig:astrophysics}, right panel) of our 1000 injections.

% I am a little bit confused, I remember the number of test set is 100, right?

\section{Microlensing surrogate parameters}

% The $l_p$ parameter is calculated according to the sum of the top 10 peaks in the time domain magnification factor.
As mentioned in the main paper ``Data and Training'' section, we use surrogate parameters $l_p$, $l_{d10}$, $l_{d60}$, and $l_{d120}$ for parametrization, as shown in the main paper FIG.~\ref{fig:parameters}. The parameter $l_p$ is calculated from the sum of the 10 highest peaks of the time-domain magnification factor, $\tilde{F}(t, \boldsymbol{y})$, defined as:
\begin{equation}
\label{eq:TimeDomainMag}
\tilde{F}\left(t, \boldsymbol{y}\right) =\int_{-\infty}^{\infty} \mathrm{d}^{2} x \delta\left[t(\boldsymbol{x}, \boldsymbol{y})-t\right]=\frac{|\mathrm{d} S|}{\mathrm{d} t } \ .
\end{equation}
Here, $l_p$ is defined relative to the macro-lensing baseline,
\begin{equation}
l_p=\sum_{i=0}^{10}F(t)_{\mathrm{peak},i}-F(t)_{\mathrm{macro}},
\end{equation}
where $F(t)_{\mathrm{macro}}=2\pi\sqrt{\mu}/k$ and $\mu$ is the strong-lensing magnification. Therefore, $l_p$ can occasionally take negative values when the summed peak contribution is below the baseline level.
In the context of microlensing, each peak in the time-domain magnification corresponds to a distinct micro-image. The physical meaning of a peak's height can be understood by analogy to the strong lensing effect near a saddle-point strong lensing image. In such cases, the strong-lensing potential is locally expanded around the strong-lensing images using the convergence $\kappa$ and shear $\gamma$. The resulting time-domain magnification factor is $\tilde{F}(t) = -2\frac{\sqrt{\mu}}{k}\ln(|t|) + C$ (see Refs.~\citep{Shan:2022xfx, Villarrubia-Rojo:2024xcj} for reference), where $\mu$ is the strong-lensing magnification, $k=4G\mathrm{M_{L}(1+z_\mathrm{L}})/c^3$, and C is a constant from numerical truncation that does not affect the frequency-domain magnification factor.

This relationship shows that the amplitude of the peak in $\tilde{F}$ is proportional to the square root of the image magnification. Applying this analogy to microlensing, we can expand the total potential around each micro-image. The magnification factor for the $i$-th micro-image is then $\tilde{F}(t)_i \approx -2\frac{\sqrt{\mu_i}}{k}\ln(|t-t_i|) + C_i$~\citep{Shan:2022xfx}, where $\mu_i$ is the magnification of that image. Therefore, the 10 highest peaks selected from $\tilde{F}(t, \boldsymbol{y})$ correspond to the 10 most magnified micro-images.

Another surrogate parameter, $l_{d[M]}$, is defined as:
\begin{equation}
\label{eq:match}
l_{d[M]} = \Im \left[ \frac{\langle h_1 \mid h_1 - h_2 \rangle}{\sqrt{\langle h_1 \mid h_1 \rangle \langle h_1 - h_2 \mid h_1 - h_2 \rangle}} \right],
\end{equation}
where $\Im$ denotes the imaginary part of the complex quantity. The term $\langle \cdot \mid \cdot \rangle$ represents the noise-weighted inner product:
\begin{align}
\langle h_1 \mid h_2 \rangle &= 4 \int_{f_{\text{low}}}^{f_{\text{high}}} \mathrm{d} f \, \frac{h_1^*(f) h_2(f)}{S_{\text{n}}(f)} \,,
\end{align}
where $S_{\text{n}}(f)$ is the single-sided power spectral density of the detector noise and $^*$ denotes the complex conjugate. In this context, $h_1$ is the unlensed waveform ($h_{\text{U}}$), and $h_2$ is the lensed waveform ($h_{\mathrm{L}}$), which includes both macro- and microlensing effects. Importantly, $l_{d[M]}$ is not a phase itself, but a signed, dimensionless mismatch-like statistic. We refer to it as ``phase-based'' because the imaginary part of the normalized inner product is sensitive to the relative phase distortion induced by microlensing: positive and negative values encode opposite signed deviations, while values near zero correspond to negligible phase distortion.

We introduce the phase-deviation parameter $l_{d[M]}$ to replace this conventional mismatch,
$%\mathrm{MM}(h_{\mathrm{U}},h_{\mathrm{L}})=
1-\frac{|\langle h_{\mathrm{U}}\mid h_{\mathrm{L}}\rangle|}
{\sqrt{\langle h_{\mathrm{U}}\mid h_{\mathrm{U}}\rangle\,\langle h_{\mathrm{L}}\mid h_{\mathrm{L}}\rangle}}$,
for two main reasons. First, the conventional mismatch is always positive, which results in a one-sided probability distribution. Training a network on such a skewed distribution can introduce bias into the parameter estimation. In contrast, $l_{d[M]}$ is distributed around zero, with zero corresponding to the null case (no microlensing), as shown in FIG.~\ref{fig:lp_ld_distrib}. This symmetric property improves performance in both parameter estimation and microlensing event selection. Second, this new parameter effectively quantifies the phase deviation induced by microlensing. A stronger microlensing effect leads to a greater phase difference between the lensed and unlensed waveforms. This, in turn, increases the value of $\Im[\langle h_1 \mid h_1 - h_2 \rangle]$, resulting in a larger value for $l_{d[M]}$. 

In this work, we employ three such deviation parameters, $l_{d10}$, $l_{d60}$, and $l_{d120}$, corresponding to binary black hole systems with component masses of $(10\,M_{\odot},\,10\,M_{\odot})$, $(60\,M_{\odot},\,60\,M_{\odot})$, and $(120\,M_{\odot},\,120\,M_{\odot})$, respectively, where each pair denotes the masses of the primary and secondary black holes, and $M_{\odot}$ is the solar mass. These parameters are designed to capture the phase-based waveform deviation across low-, intermediate-, and high-mass binary black hole events, enabling a more comprehensive characterization of microlensing effects over a broad mass range.
%In addition to $l_{d[M]}$, we analyze the time-domain peak parameter $l_p$. Unlike the phase-based $l_{d[M]}$, $l_p$ shows a positive bias in FIG.~\ref{fig:lp_ld_distrib}. It is worth noting that $l_p$ can occasionally be negative, as it is defined relative to a baseline value (see the dashed line in Fig.~\ref{fig:parameters} of the main paper), namely $l_p = \sum_{i=0}^{10} F(t)_{\mathrm{peak},i} - F(t)_{\mathrm{macro}}$, where $F(t)_{\mathrm{macro}}$ is defined as $2\pi\sqrt{\mu}/k$ and $\mu$ is the strong-lensing magnification. 
The combination of the symmetric phase parameters $l_{d[M]}$ and the amplitude-sensitive parameter $l_p$ contributes to better-calibrated posteriors in flow-based inference; this calibration is quantified later with a Probability--Probability (P--P) plot.

% \subsection{Microlensing surrogate parameter distribution in the training dataset}

% We present the distribution of the lensing parameters, the peak value $l_p$, and deviation parameters $l_{d[M]}$, used in the training dataset. As shown in FIG.~\ref{fig:lp_ld_distrib}, all parameters exhibit distributions peaked around zero, consistent with the expectation that most gravitational-wave microlensing events induce only subtle distortions. The peak parameter in the time domain, $l_p$, shows a positive bias, which reflects its definition as the sum of peak values in the time domain. It is worth mentioning that the peak parameter $l_p$ can be less than zero. This is due to the definition of $l_p$, which is the peak value in FIG.~\ref{fig:parameters} minus the baseline (black dashed line) value. In some cases, the peak is below the dashed line. In contrast, the deviation parameters $l_{d[M]}$ are centered around zero and display approximately symmetric distributions. This symmetry arises from the phase-based nature of the $l_{d[M]}$ definition, which allows for both positive and negative fluctuations. In practice, these symmetric distributions contribute to well-calibrated $p$-values in flow-based posterior inference. 

\section{Normalizing flow and training}

%Given a micro-lensed waveform $h_{lensed}$, the goal of our model is to perform the density estimation $q(\theta_{\rm GR}, \theta _s|h_{lensed})$ of both BBH gravitational wave parameters $\theta_{\rm GR}$ and lensing parameters $\theta _s$. Following~\cite{dingo2021}, we train a conditional normalizing flow – specifically a neural spline flows (NSF)~\cite{nsf_durkan2019neural} that models complex, potentially multimodal posteriors through adaptive spline transformations – which parameterizes a flexible probability distribution over the joint space of astrophysical and lensing parameters. Specifically, the flow learns an invertible transformation $f_\phi$ that maps samples from a simple base distribution (e.g., a multivariate Gaussian) to samples from the complex, high-dimensional posterior distribution, conditioned on the observed lensed waveform data $h_{lensed}$. The parameters of the flow are optimized during training to maximize the likelihood of the training data under the model, effectively learning the density estimation of the full joint posterior distribution, which is critical for simultaneously inferring both the properties of the BBH system and the characteristics of the gravitational microlens.
Given a microlensed waveform $h_\mathrm{L}$, the goal of our model is to estimate the joint posterior $q(\theta_{\rm GR}, \theta _s|h_\mathrm{L})$ of both the BBH gravitational wave parameters $\theta_{\rm GR}$ and the lensing parameters $\theta _s$. Following~\citet{dingo2021}, we train a conditional normalizing flow, specifically a neural spline flow (NSF)~\citep{nsf_durkan2019neural} that models complex, potentially multimodal posteriors via adaptive spline transformations. 
The flow learns an invertible transformation $f_\phi$ that maps samples from a simple base distribution (e.g., a multivariate Gaussian) to samples from the complex, high-dimensional posterior distribution, conditioned on the observed lensed waveform data $h_\mathrm{L}$. 
The network parameters of the flow are optimized during training to maximize the likelihood of the training data under the model, effectively learning the joint posterior over both the BBH and microlensing parameters.
In each training epoch, the BBH signals are shuffled and used once per epoch, while microlensed signals are randomly paired with BBH signals under the redshift constraint $z_{\rm L}\le z_s$ (consistent with the training-set sampling ranges above). Each microlensing waveform is reused ten times on average per epoch.

\subsection{Normalizing-flow Validation with a Simple Non-Deterministic Waveform}

\begin{figure*}
    \centering
    \includegraphics[trim={150pt 150pt 100pt 100pt},clip,width=\linewidth]{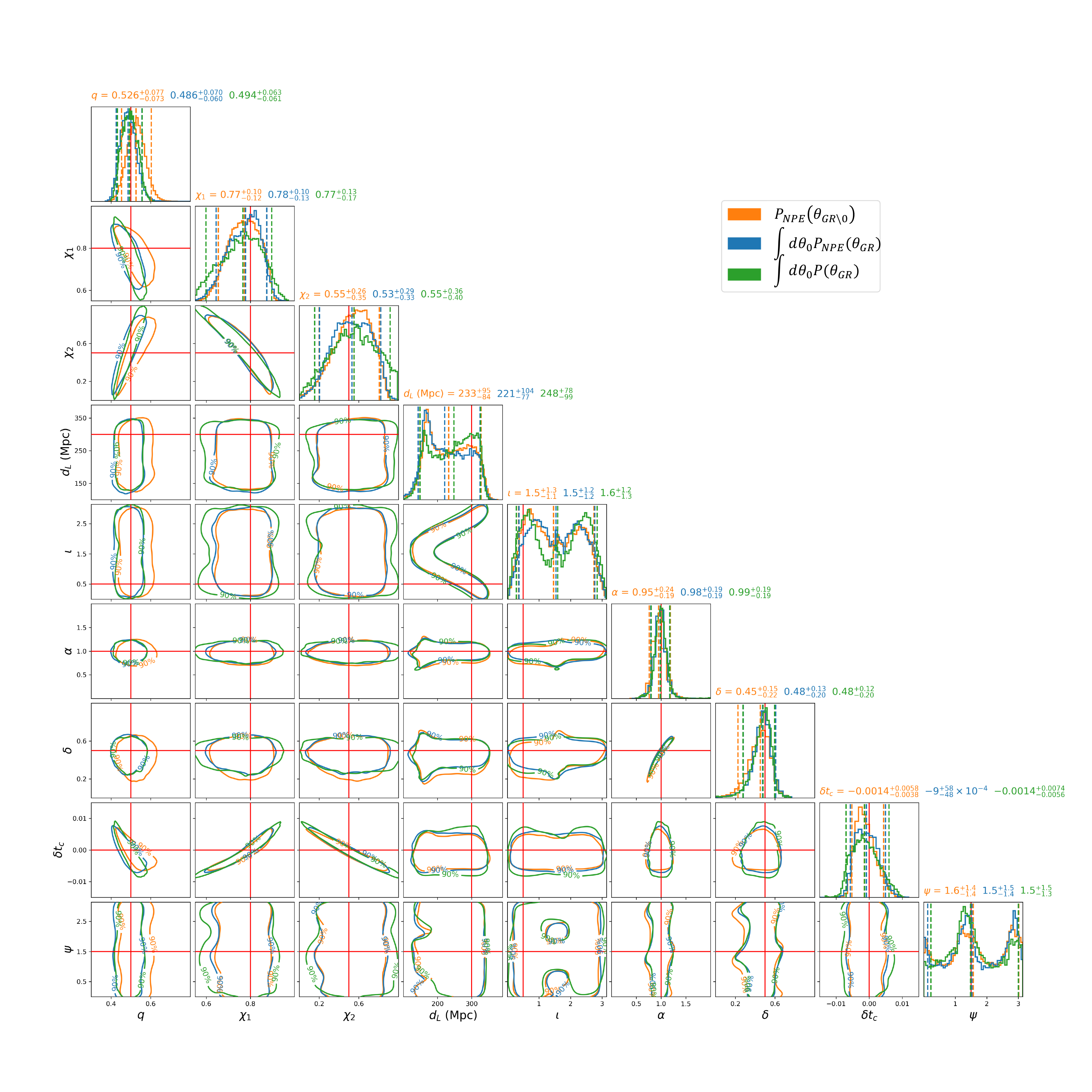}
    \caption{The posterior distributions of the mass ratio $q$, aligned spins of the primary and secondary BHs ($\chi_1$ and $\chi_2$, respectively), luminosity distance $d_\mathrm{L}$, inclination angle $\iota$, right ascension $\alpha$, declination $\delta$, reference merger time $\delta t_c$, and polarization $\psi$ are shown. The orange curve labeled ``$P_{\rm NPE}(\theta_{\rm GR\setminus 0})$'' shows the Neural Posterior Estimation (NPE) result without including chirp mass as an inference parameter. The blue curve labeled``$\int d \theta_0 P_{\rm NPE}(\theta_{\rm GR})$'' represents the NPE result with chirp mass included as an inference parameter. The green curve labeled``$\int d \theta_0 P(\theta_{\rm GR})$'' corresponds to the result obtained from a traditional MCMC simulation using the {\it emcee\_pt} sampler embedded in the PyCBC package~\cite{alex_nitz_2022_6324278}. The solid curves in the 2D contour plots represent the 90\% credible regions, while the vertical dashed lines indicate the two-sided 90\% confidence intervals. The solid red lines mark the injected (true) parameter values.}
    \label{fig:npe_vs_MC}
\end{figure*}

\begin{figure*}
    \centering
    \includegraphics[width=\linewidth]{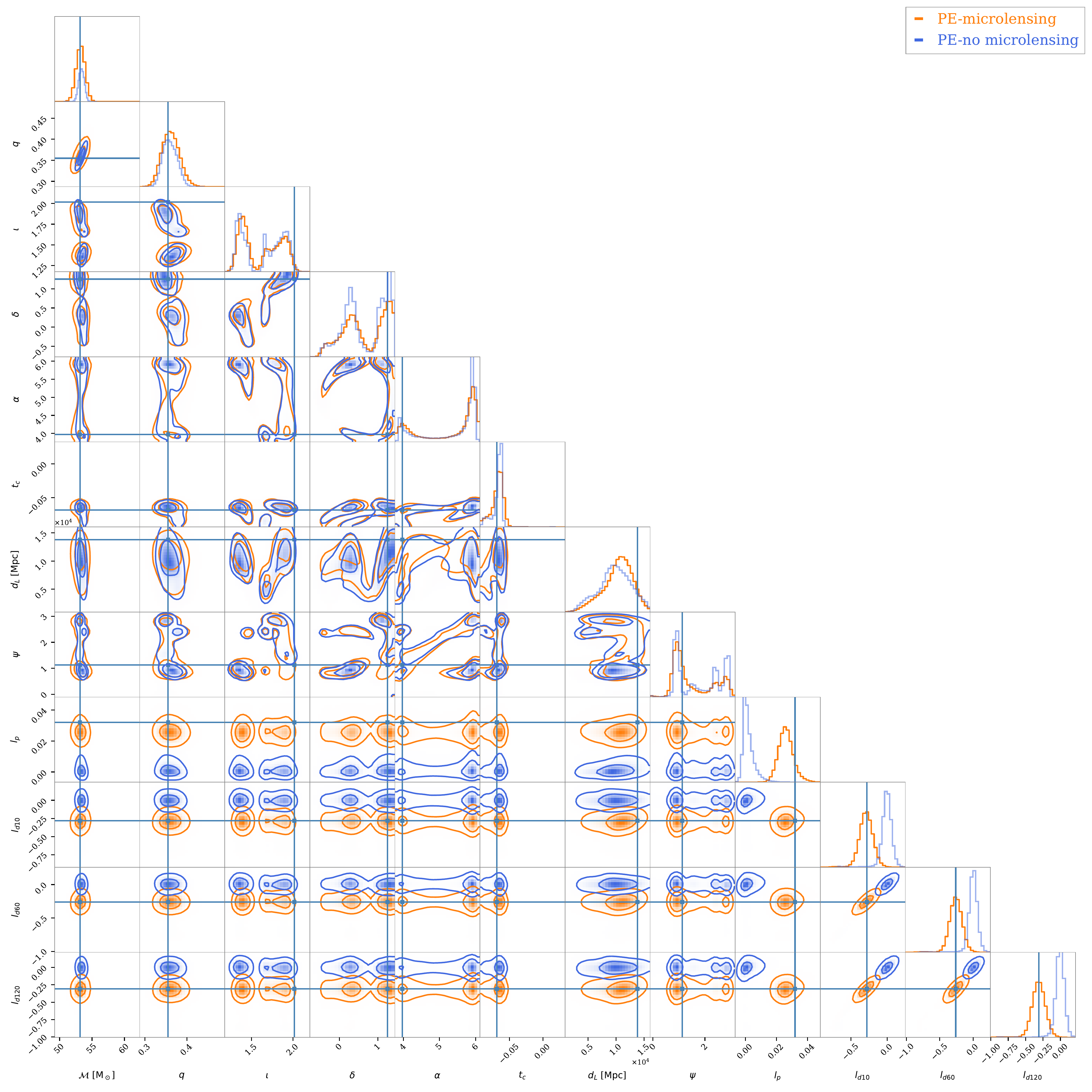}
    \caption{A sampled parameter estimation (PE) result from microlensing and null-microlensing injection, for all parameters. Microlensing parameters are $l_p$ and $l_{d[M]}$ illustrated in ``Microlensing surrogate parameters'' section, while BBH parameters includes chirp mass $\mathcal{M}, $mass ratio $q$, inclination angle $\iota$, declination $\delta$, right ascension $\alpha$, time of coalescence at geocenter $t_c$, luminosity distance $d_\mathrm{L}$ and polarization angle $\psi$.}
    \label{fig:PE_total}
\end{figure*}

Inspired by recent works on neural posterior estimation and marginalization in gravitational-wave and cosmological inference \citep{vanstraalen2024premergerdetectioncharacterizationinspiraling,PhysRevD.109.123547}, we examine whether a normalizing-flow framework can consistently recover marginalized posterior distributions of selected observables in a non-deterministic waveform setting.
To validate this idea, 
%We argue that the normalizing-flow framework naturally yields posterior distributions for selected characteristic observables of a nondeterministic waveform. To validate this approach, 
we test this proposal with a simple example using the binary black hole waveform. We consider the BBH waveform with GR parameter $\theta_{\rm GR}$ and split it into $(\theta_0,\theta_{\rm GR\setminus 0})$. Here, $\theta_0$ is chosen as the chirp-mass parameter used in the BBH waveform model. The chirp mass is $\mathcal{M}_c=(m_1m_2)^{3/5}(m_1+m_2)^{-1/5}$ and is distinct from an individual component mass. Source-frame and detector-frame chirp masses are related by $\mathcal{M}_c^{\rm det}=(1+z)\mathcal{M}_c^{\rm src}$. In this proof-of-principle test, $\theta_0$ is introduced only as the parameter to be marginalized over and is treated as $\theta_{*\setminus s}$, which is not explicitly labeled in neural network training. The normalizing flow then yields the posterior distribution of $P_{\rm NPE}(\theta_{\rm GR\setminus 0})$, which can be validated by checking whether it satisfies
\begin{align}
P_{\rm NPE}(\theta_{\rm GR\setminus 0}) \approx P(\theta_{\rm GR\setminus 0}) =\int d \theta_0 P(\theta_{\rm GR})\,.
\end{align}
Notice that in this example $P(\theta_{\rm GR})$ is known through Monte-Carlo methods, so comparison can be explicitly performed.

FIG.~\ref{fig:npe_vs_MC} compares posterior distributions for key BBH parameters using Neural Posterior Estimation (NPE) with and without chirp mass, and traditional Monte-Carlo Markov Chain (MCMC) via PyCBC~\citep{alex_nitz_2022_6324278}.
The posterior distribution $P_{\rm NPE}(\theta_{\rm GR\setminus 0})$ is largely consistent with both $\int d \theta_0 P(\theta_{\rm GR})$ and $\int d \theta_0 P_{\rm NPE}(\theta_{\rm GR})$, where the training of $P_{\rm NPE}(\theta_{\rm GR})$ corresponds to deterministic waveforms. 

For a quantitative comparison, we evaluate the Jensen--Shannon divergence (JSD) between the marginalized posteriors for each of the nine parameters for every pair among the three distributions
$P_{\rm NPE}(\theta_{\rm GR\setminus 0})$,
$\int d\theta_0\,P_{\rm NPE}(\theta_{\rm GR})$,
and
$\int d\theta_0\,P(\theta_{\rm GR})$.
These correspond, respectively, to NPE trained without labeling the chirp mass, NPE trained on the full parameter set and then marginalized over the chirp mass, and the MCMC reference marginalized over the chirp mass.
As a symmetric divergence, the JSD ranges from $0$ to $\ln 2 \approx 0.69$~nat, with smaller values indicating closer agreement~\cite{lin2002divergence,romero2020bayesian}. We find that, for any pairwise comparison among the three posteriors, all JSD values remain below $0.1$, with most values around $0.01$, and the largest discrepancy occurring for $q$ ($\mathrm{JSD}=0.075$), while the remaining parameters are even smaller. These results indicate that all three posteriors are highly consistent when compared at the level of their individual marginal distributions.
Although this promising consistency has only been verified in a specific scenario, it suggests that a similar relation may still hold for more general settings, which motivates the application of normalizing flow to infer the posterior distribution of lensing parameters.

%Focusing on the important scenario of stellar-field microlensing, we take an initial step in this direction with neural-network techniques. We argue that the normalizing-flow framework naturally yields posterior distributions for selected characteristic observables of a nondeterministic waveform. To validate this approach, we test this proposal with a simple example using the binary black hole waveform and treating the chirp mass as a latent variable (see Supplementary FIG.~1). In this example, where Monte Carlo–based parameter inference remains tractable, the nondeterministic inference using normalizing flow~\citep{nsf_durkan2019neural} produces consistent results as the Monte-Carlo method. This finding motivates us to further apply the same technique for more complex microlensed waveforms.

%Here, it is worth mentioning that the peak parameter $l_p$ can be less than zero. This is due to the definition of $l_p$, which is the peak value in FIG.~\ref{fig:parameters} minus the baseline (black dashed line) value. In some cases, the peak is below the dashed line, making $l_p$ less than zero.

\subsection{Parameter Estimation result for all predicted parameters}

%We show a typical PE result of the injection in Fig.~\ref{fig:PE_total}, with all predicted parameters. Similar to Fig.~\ref{fig:PE_lens_mah_redshift}, the blue and red colors separately indicate the micro-lensing and the corresponding strong-lensing signal. Both injections yield consistent posterior distributions for the intrinsic source parameters, as expected for GW signals originating from the same binary system, while their lensing-related parameters exhibit different distributions. The result validates our framework’s ability to disentangle these two lensing scenarios.
We show a typical PE result in FIG.~\ref{fig:PE_total}, with all predicted parameters, including BBH parameters and microlensing parameters. The orange and blue distributions correspond to the microlensed and the associated non-microlensed signals, respectively, as illustrated in the main paper ``Parameter Estimation results and implications'' section. Both injections yield consistent posterior distributions for the BBH parameters, as expected for gravitational-wave signals originating from the same binary system. In contrast, the microlensing surrogate parameters exhibit distinct distributions (same as main paper FIG.~\ref{fig:PE_lens_mah_redshift}), reflecting differences in their lensing configurations. This result demonstrates the capability of our framework to disentangle microlensing effects from strong lensing in a coherent and physically interpretable manner.

\begin{figure}
    \centering
    \includegraphics[width=\linewidth]{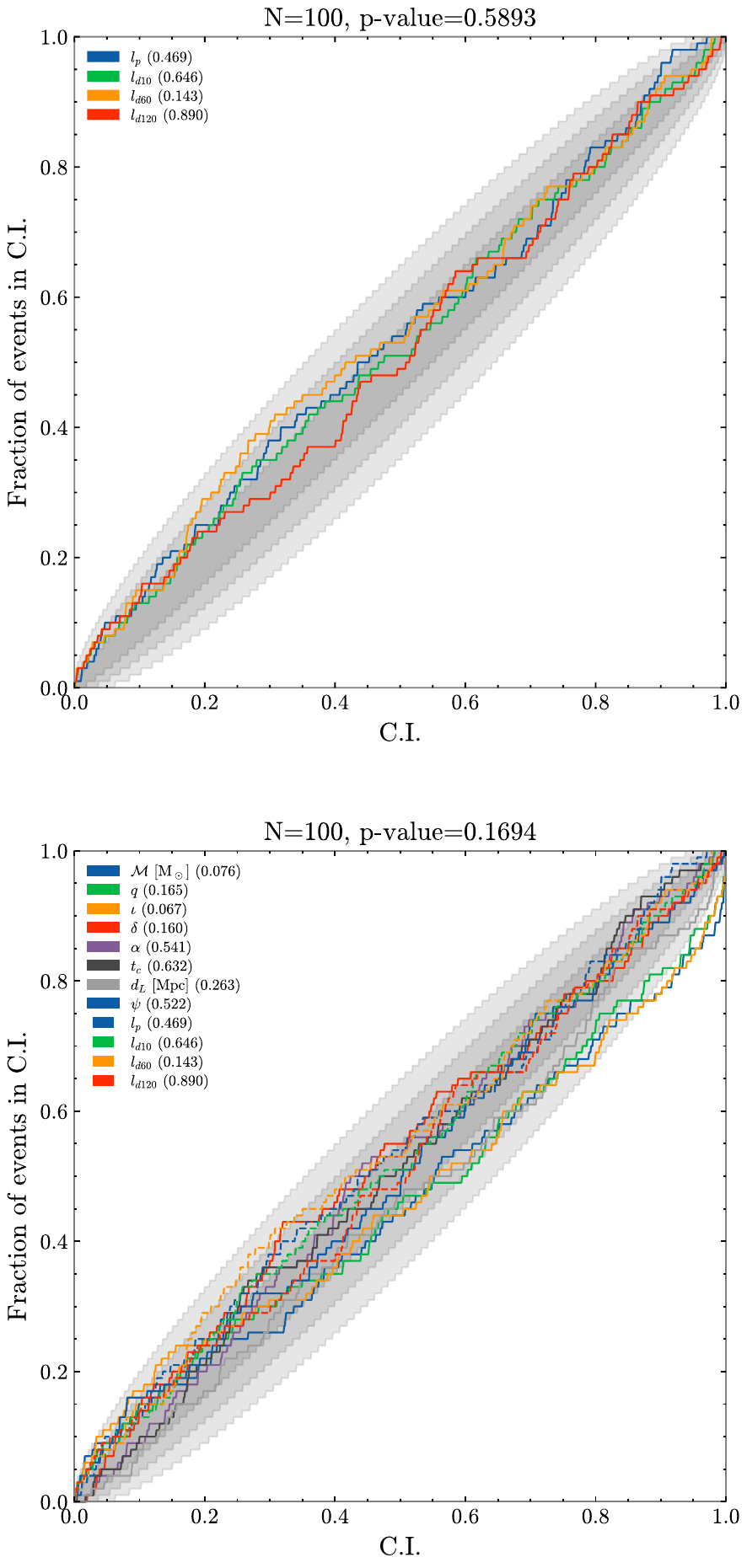}
    \caption{Probability-Probability (P-P) plot for lensing parameters (top) and all parameters (bottom). The x-axis denotes the credible interval (C.I.), and the y-axis shows the fraction of injected events whose true parameter values lie within that interval. A well-calibrated posterior follows the diagonal line, indicating consistency between the predicted credible intervals and their empirical coverage. The numbers in parentheses in the legend are the Kolmogorov--Smirnov $p$-values for the corresponding one-dimensional marginalized parameters. The shaded regions indicate the expected $1\sigma$, $2\sigma$, and $3\sigma$ deviations from the diagonal for a perfectly calibrated posterior given the finite sample size.}
    \label{fig:ppplot_total}
\end{figure}

\subsection{Probability-Probability (P-P) plot}

%To further evaluate our model's ability for lensing parameter posterior estimation, we sample posteriors from 100 injections consistent with the training distribution and draw a P-P plot for the lensing parameters, as shown in Fig.~\ref{fig:ppplot_total}. Specifically, for the true lensing parameter value, we compute its percentile position in the marginalized posterior, evaluate the CDF, and visualize the results in the diagram. Fig.~\ref{fig:ppplot_total} left shows the P-P plot for the micro-lensing parameters, while Fig.~\ref{fig:ppplot_total} right shows the P-P plot for all the inferred parameters, including BBH parameters. The Kolmogorov-Smirnov test p-values and the combined p-value (0.5893 for lensing parameters, 0.1694 for all parameters) are also shown in the figure, confirming that our model’s predictions for lensing parameters and other parameters are statistically consistent with expectations.

To further evaluate our model's ability for lensing parameter posterior estimation quantitatively, we perform a Probability-Probability (P-P) plot analysis using 100 injection events sampled consistently from the training distribution, as shown in FIG.~\ref{fig:ppplot_total}. For each event and each parameter, we determine the credible level at which the true injected value is recovered in the corresponding marginalized posterior. The P--P plot then shows the fraction of events recovered within a given credible interval; for a well-calibrated posterior this relation follows the diagonal. The top panel of FIG.~\ref{fig:ppplot_total} shows the P–P plot restricted to microlensing parameters, while the bottom panel presents the result for all inferred parameters. 

The Kolmogorov–Smirnov test yields $p$-values of $0.5893$ for the lensing parameters and $0.1694$ for the full parameter set, both suggesting no statistically significant deviation from the expected uniform distribution within the tested sample size. 
The minor fluctuations (e.g. $\mathcal{M}$) likely arise from residual degeneracies between microlensing-induced waveform distortions and intrinsic binary parameters. It is difficult to uniquely attribute such small deviations to a specific intrinsic parameter, since multiple parameters can partially compensate for subtle waveform distortions. 
%The minor fluctuations (e.g. $\mathcal{M}$) likely arise from its role as the leading-order parameter scaling the gravitational-wave amplitude. Since $\mathcal{M}$ primarily dictates the overall signal strength, it can partially absorb residual magnification effects in cases of subtle microlensing.
The phase parameters $l_{d[M]}$, whose distributions are centered about zero, and the amplitude-sensitive parameter $l_p$ improve the representation of microlensing features and lead to better-calibrated posterior estimates, although completely eliminating these degeneracies remains challenging for complex stochastic lensing signals.
%Besides, even with the incorporation of symmetry-aware parameters (e.g., $l_{d[M]}$), the network can still encounter difficulties in perfectly capturing the full posterior distribution for every sampled event.
Overall, these results confirm that the posterior estimates produced by our model are well-calibrated and statistically consistent with the true injected values across the full parameter space.

% \begin{figure}
%     \centering
%     \includegraphics[width=\linewidth]{fig/pp-plot.pdf}
%     \caption{P-P plot}
%     \label{fig:ppplot}
% \end{figure}

\subsection{Fraction of detectable microlensing events in the \texorpdfstring{$\kappa$--$\kappa_s$}{kappa--kappa-s} plane}
\label{subsec:frac_kappa_kappas}

\begin{figure}[h]
    \centering
    \includegraphics[width=0.5\textwidth]{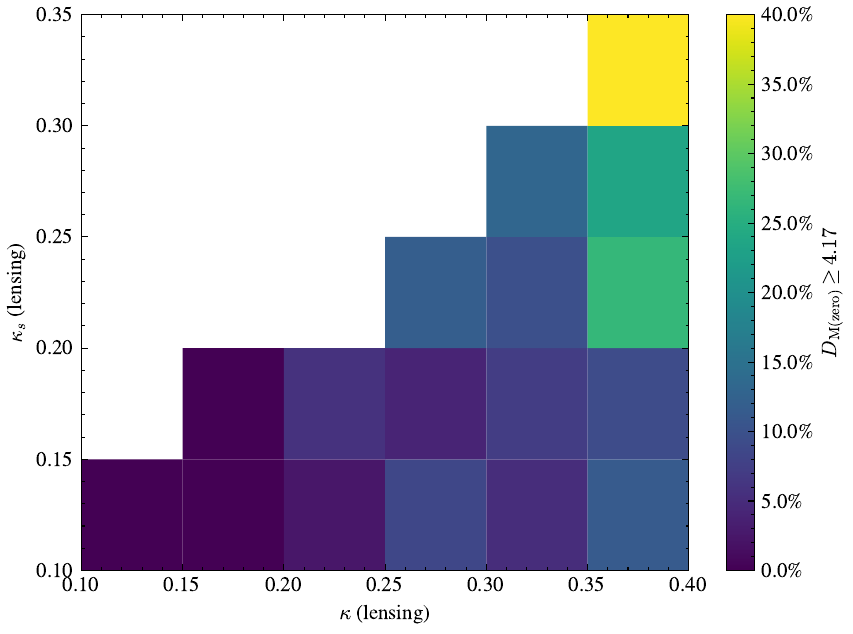}
    \caption{\textbf{Fraction of detectable microlensing events in the $\kappa$–$\kappa_s$ plane.} The color map shows the fraction of simulated events with a Mahalanobis distance $D_{\mathrm{M}} \ge 4.17$ ($3\sigma$ threshold derived from the null distribution), representing statistically significant microlensing detections, as a function of the total convergence $\kappa$ and the stellar convergence $\kappa_s$. Each bin in the $\kappa$–$\kappa_s$ grid has a size of 0.05, and the color scale indicates the percentage of events within that bin.}
    \label{fig:ka_kas_mah_dis_over3}
\end{figure}

As discussed in the main text, our model reveals a correlation between the Mahalanobis distance $D_{\text{M(zero)}}$ and the lensing convergence quantities $\kappa$ and $\kappa_s$. To further illustrate the relationship between the microlensing detectability and the underlying stellar field, we compute the fraction of simulated events with a Mahalanobis distance $D_{\mathrm{M(zero)}} \ge 4.17$ ($3\sigma$ threshold derived from the null distribution) across the $\kappa$–$\kappa_s$ parameter plane. As shown in FIG.~\ref{fig:ka_kas_mah_dis_over3}, each bin in the $\kappa$–$\kappa_s$ grid (bin size of 0.05) represents the percentage of events that reach the $3\sigma$ confidence level for microlensing detection. The distribution shows a clear increasing trend with both $\kappa$ and $\kappa_s$, indicating that microlensing effects become more prominent in regions of higher total convergence and denser stellar fields. This result provides additional support for the physical correlation between our microlensing discriminators and the lens galaxy’s stellar population.

\subsection{Impact of redshift and chirp mass on microlensing detectability}

\begin{figure}[h]
    \centering
    \includegraphics[width=0.5\textwidth]{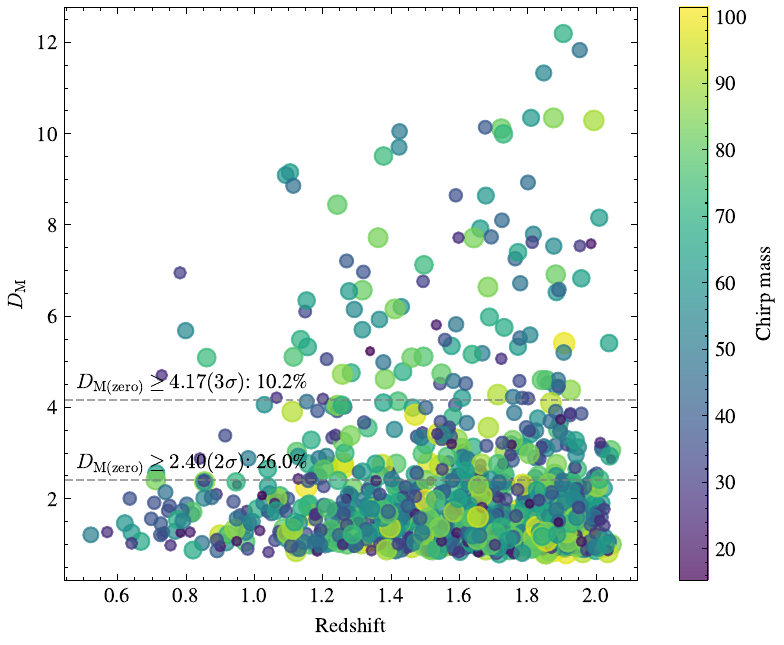}
    \caption{Mahalanobis distance distribution with respect to the redshift and the chirp mass of the BBH.}
    \label{fig:Mah_redshift}
\end{figure}

Microlensing might be expected to be most detectable in low-redshift (high-SNR) or low-chirp-mass events, where waveform distortions are stronger at high frequencies. However, FIG.~\ref{fig:Mah_redshift}, which shows the Mahalanobis distance as a function of redshift and chirp mass, reveals no strong preference for either parameter. This indicates that detectability is not restricted to a narrow subset of the astrophysical population.

This can be understood from the approximate scaling $\langle H-h \mid H-h \rangle \approx \rho^2 MM$, where $H$ and $h$ denote the lensed and unlensed GW signals, respectively, $\rho$ is the GW signal-to-noise ratio (SNR), and $MM$ is the microlensing-induced mismatch. Low-redshift events usually have higher SNRs, but they also tend to have smaller effective microlensing masses because the lensing effect scales with $(1+z_\mathrm{L})$, where $z_\mathrm{L}$ is the lens redshift. 
Likewise, lower-chirp-mass events generally provide more waveform cycles within the detector band and extend to higher frequencies, which can enhance the measurability of microlensing-induced distortions. However, they generally have lower SNRs.
%Likewise, lower-chirp-mass events produce larger microlensing mismatches because they radiate at higher frequencies, but they generally have lower SNRs. 
These competing effects weaken any strong bias toward specific redshifts or chirp masses. A more complete assessment of the redshift dependence would require extending the simulated population beyond the current $z_s\leq2$ range and jointly accounting for the SNR distribution and the redshift dependence of the microlensing effect, which we leave for future studies.

%\section{microlensing surrogate parameters}
\end{document}